\documentclass[aps,reprint,twocolumn,superscriptaddress,floatfix,preprintnumbers,showpacs]{revtex4-2}
\usepackage{graphicx,graphics,amssymb,amsmath,bbm,appendix}
\usepackage{inputenc}
\usepackage{comment}
\usepackage{hyperref}
\usepackage[dvipsnames]{xcolor}

\DeclareMathOperator{\omres}{\omega_0}
\DeclareMathOperator{\om}{\delta\omega}
\DeclareMathOperator{\gt}{\mathsf{g}}
\DeclareMathOperator{\ef}{\mathcal{E}}
\DeclareMathOperator{\ktwo}{\kappa_2}
\DeclareMathOperator{\kthree}{\kappa_3}
\DeclareMathOperator{\vzpf}{\mathnormal{V}_{\rm zpf}}
\DeclareMathOperator{\fid}{\mathcal{F}}
\DeclareMathOperator{\vg}{\mathnormal{V}_\mathnormal{G}^{0}}
\DeclareMathOperator{\kr}{\mathnormal{\kappa_r}}
\DeclareMathOperator{\keff}{\mathnormal{\kappa_{\rm eff}}}

\newcommand{\unit}[1]{\,{\rm #1}}
\newcommand{\av}[1]{{\rm E}[#1]}

\newcommand{\mf}[1]{#1}

\begin{document}

\title{Tunable hole spin-photon interaction based on $\gt$-matrix modulation}
\author{V. P. Michal}
\author{J. C. Abadillo-Uriel}
\affiliation{Univ. Grenoble Alpes, CEA, IRIG-MEM-L\_Sim, Grenoble, France.}
\author{S. Zihlmann}
\author{R. Maurand}
\affiliation{Univ. Grenoble Alpes, CEA, Grenoble INP, IRIG-Pheliqs, Grenoble, France.}
\author{Y.-M. Niquet}
%\email{yniquet@cea.fr}
\author{M. Filippone}
\email{yniquet@cea.fr, michele.filippone@cea.fr}
\affiliation{Univ. Grenoble Alpes, CEA, IRIG-MEM-L\_Sim, Grenoble, France.}

\begin{abstract}
We consider a spin circuit-QED device where a superconducting microwave resonator is capacitively coupled to a single hole confined in a semiconductor quantum dot. Thanks to the strong spin-orbit coupling intrinsic to valence-band states, the gyromagnetic $\gt$-matrix of the hole can be modulated electrically. This modulation couples the photons in the resonator to the hole spin. We show that the applied gate voltages and the magnetic-field orientation enable a versatile control of the spin-photon interaction, whose character can be switched from fully transverse to fully longitudinal. The longitudinal coupling is actually maximal when the transverse one vanishes and vice-versa. This ``reciprocal sweetness'' results from geometrical properties of the $\gt$-matrix and protects the spin against dephasing or relaxation.  
We estimate coupling rates reaching $\sim 10\unit{MHz}$ in realistic settings and discuss potential circuit-QED applications harnessing either the transverse or the longitudinal spin-photon interaction. Furthermore, we demonstrate that the $\gt$-matrix curvature can be used to achieve parametric longitudinal coupling with enhanced coherence.
\end{abstract}

\maketitle

\twocolumngrid

%%%%%%%%%%%%%%%%%%%%%%%%%%%%%%%
%   INTRO
%%%%%%%%%%%%%%%%%%%%%%%%%%%%%%%

Hybrid circuit quantum electrodynamics (cQED) investigates the interaction between microwave photons in resonators and diverse quantum excitations in solid state devices~\cite{Clerk2020}. This field inherits from cavity QED~\cite{haroche2006exploring} its fundamental interest in light-matter interactions, and sustains the development of quantum information~\cite{Blais2020} by extending the research initiated with superconducting circuits~\cite{Blais2021}. Special efforts are currently devoted to the coupling of microwave photons with the charge~\cite{mi2017strong,Stockklauser2017,bruhat_2018_cQED,scarlino2021situ} and spin~\cite{Imamoglu1999,Viennot2015,Samkharadze2018,Mi2018,Landig2018,Yu2022} degrees of freedom of a localized electron or hole. The spin of a single carrier is indeed of particular interest as it is one of the most elemental realizations of a two-level system~\cite{Burkard2021}.

Light acts on two-level systems
either {\it transversally} or {\it longitudinally}, that is by changing or preserving the level occupations, respectively. 
\mf{Transverse coupling underpins high-ﬁdelity dispersive readout~\cite{Walraff2004} and control operations~\cite{sandbergTuningFieldMicrowave2008,delbecq2013photon,sandbergTuningFieldMicrowave2008,yin_catch_and_release_2013,Zajac2018,wangSchrodingerCatLiving2016,ofekExtendingLifetimeQuantum2016}, while longitudinal coupling has recently emerged as a valuable resource for fast quantum non-demolition measurements~\cite{Didier2015,Richer2016,Richer2017,PhysRevLett.122.080503,PhysRevLett.122.080502,dasonneville_PRX_2020} and multi-qubit entangling gates~\cite{PhysRevA.86.032324,roos2008ion, Jin2012, Kerman2013, Billangeon2015, Richer2016, Royer2017, Schuetz2017, Schuetz2019, Harvey2018, Bottcher2021, Ruskov2019, Ruskov2021}.}  
Yet, controlling the relative strength of longitudinal and transverse couplings requires specific circuit design \cite{Didier2015,Richer2016,Richer2017,PhysRevLett.122.080503,PhysRevLett.122.080502,dasonneville_PRX_2020}, driving protocols~\cite{lambert2018amplified} or additional elements such as micro-magnets~\citep{abadillo2019enhancing}. \mf{Relying on intrinsic 
physical mechanisms to switch from one coupling to the other may thus open new opportunities to explore and harness various aspects of light-matter interaction}.

\begin{figure}[t]
\centering
\includegraphics[width=\columnwidth]{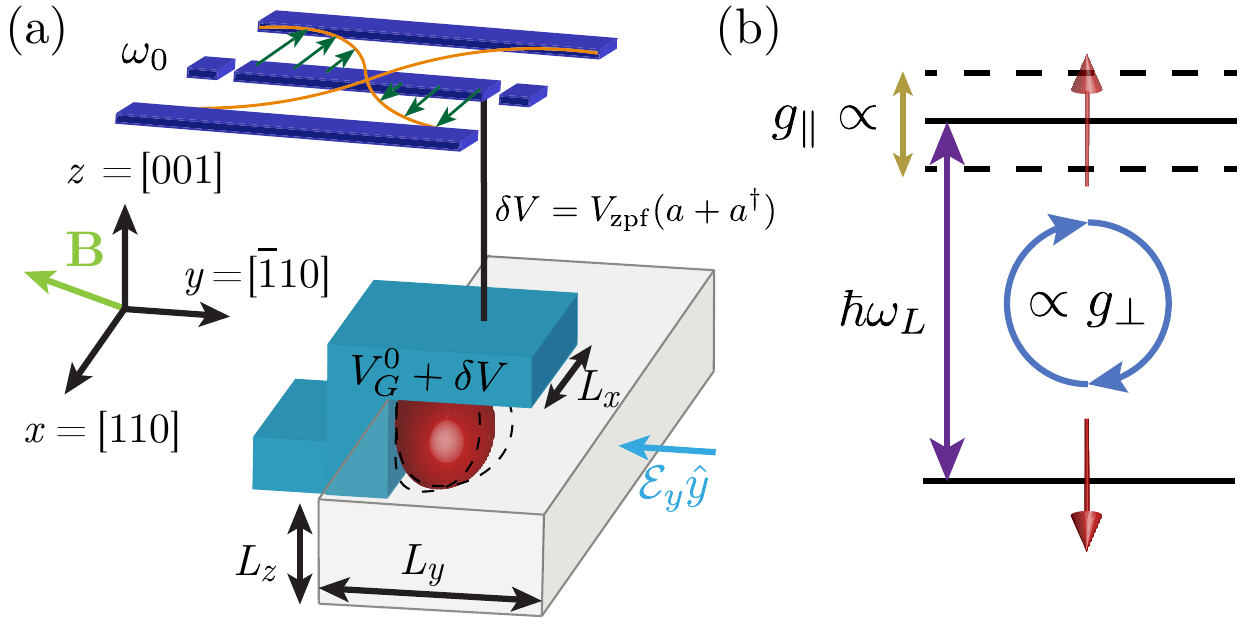}
    \caption{(a) Sketch of the setup considered here. A single hole quantum dot (red shape) is confined in a rectangular nanowire with sides $L_y>L_z$ by a partly overlapping gate (light blue) with length $L_x$ and DC bias $V=\vg$. The hole is coupled to photons in a microwave coplanar resonator connected to the gate. In this configuration, the shape of the quantum dot (dashed lines) is primarily controlled by the in-plane component $\ef_y$ of the electric field of the gate \cite{Venitucci2018,Venitucci2019,Michal2021}, which modulates the Larmor vector $\boldsymbol\omega_L$ of the hole spin owing to spin-orbit coupling. (b) As a consequence, quantized voltage fluctuations $\delta V$ originating from the resonator photon field can change the norm of $\boldsymbol\omega_L$ (longitudinal coupling $g_\parallel$) and/or its orientation (spin flip transverse coupling $g_\perp$).}
    \label{fig:setup}
\end{figure}

In this work, we show that a single hole confined in a compact semiconductor nanostructure exhibits a controllable spin-photon interaction, without the need for additional driving or circuit elements \citep{Mi2018, Samkharadze2018}. In such a spin cQED system, the key to tunability is the presence of an {\it intrinsically} strong spin-orbit coupling (SOC) allowing for all-electrical modulation of the hole gyromagnetic $\gt$-matrix ~\cite{Kato2003, Ares2013,Venitucci2018, Venitucci2019, Michal2021}. This strong SOC underpins hole-spin qubit control in silicon~\cite{Maurand2016,Crippa2018,Camenzind2022} and germanium~\cite{Watzinger2018, Hendrickx2020_single_hole}, as well as multi-qubit logic~\cite{Hendrickx2020_two_qubits,Hendrickx2021}. Thanks to SOC, photons in a microwave resonator directly couple to the hole spin. With the minimal device depicted in Fig.~\ref{fig:setup}a, we demonstrate that the gate voltages and the orientation of the applied magnetic field provide flexible control over the spin-photon interaction. In particular, we find that the spin-photon coupling can be switched from fully transverse to fully longitudinal owing to geometrical properties of the hole spin $\gt$-matrix. Moreover, we show that strong parametric longitudinal coupling with long coherence can also be achieved. With realistic parameter estimates we discuss potential applications such as quantum-state transfer and CZ gates. Our work thus highlights the potential of hole quantum dots in semiconductors as compact and versatile platforms for cQED experiments with spins.

%%%%%%%%%%%%%%%%%%%%%%%%%%%%%%%
%   G-matrix MODULATION
%%%%%%%%%%%%%%%%%%%%%%%%%%%%%%%

We start by discussing the physics of the hole qubit device sketched in Fig.~\ref{fig:setup}, and outline how it features all the ingredients needed to achieve tunable spin-photon coupling. \mf{The choice of this device is motivated by its simplicity, where a minimal set of elements -- one gate and one resonator capacitively coupled to a semiconducting slab -- features different and tunable spin-photon interactions. However, the considerations here exposed are general and readily extend to arbitrary and more complex device geometries. }
It is well established~\cite{Kato2003,Ares2013,BraunsPRBelectric,Crippa2018,Venitucci2018,Venitucci2019,Michal2021,LilesPRB2021,Piot2022} that strong SOC in the valence band of semiconductors leads to an anisotropic Zeeman interaction between the spin of a single hole and an external magnetic field $\mathbf B$. This interaction is generally described by a two-level Hamiltonian
\begin{equation}\label{eq:spin}
\mathcal H =\frac{\mu_B}2\,\boldsymbol\sigma\cdot\gt(V)\cdot\mathbf B\,,
\end{equation}
where $\boldsymbol\sigma$ is the vector collecting the Pauli matrices of the hole spin, $\mu_B$ is the Bohr magneton, and $\gt(V)$ is the $\gt$-matrix~\cite{Abragam1970,Crippa2018,Venitucci2018}.  
Differently from electrons in vacuum, the $\gt$-matrix is not a number ($\gt_e\simeq2$) and it strongly depends on the voltage $V$ applied to confine the hole~\footnote{\mf{In principle, the interactions of electrons confined in quantum dots with a magnetic field are also described by a $\gt$-matrix. Nevertheless, the deviations from a scalar $\gt$-factor are much weaker and often neglected, especially in Silicon and Germanium~\cite{winkler2003spin}.}}.

\begin{figure*}[t]
\centering
  \includegraphics[width=0.32\textwidth]{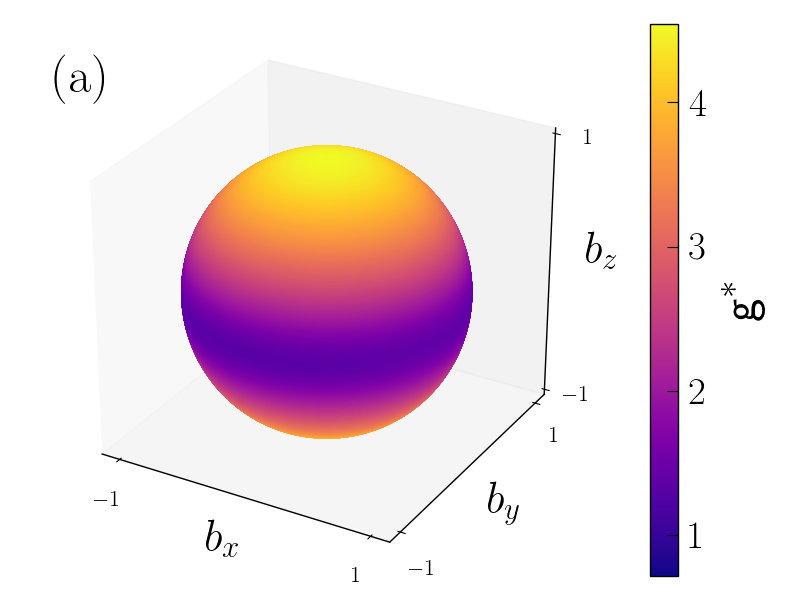}
  \includegraphics[width=0.32\textwidth]{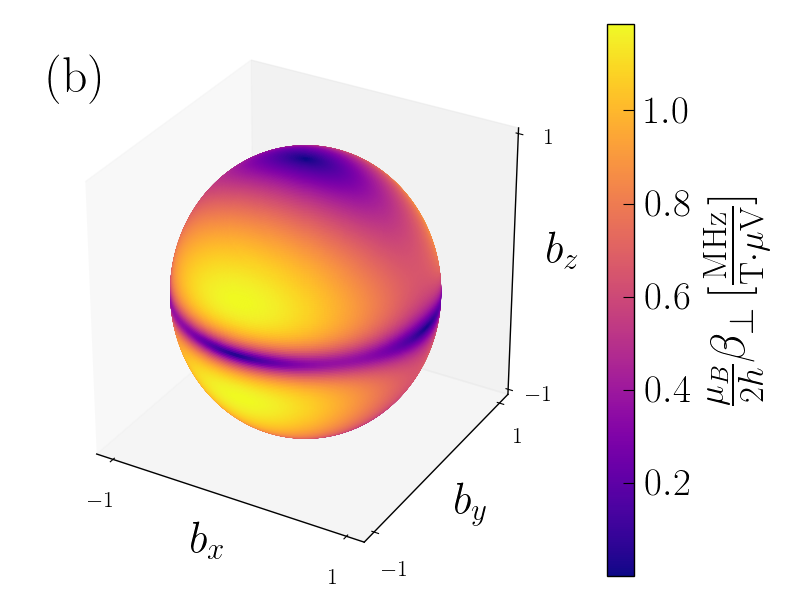}
  \includegraphics[width=0.32\textwidth]{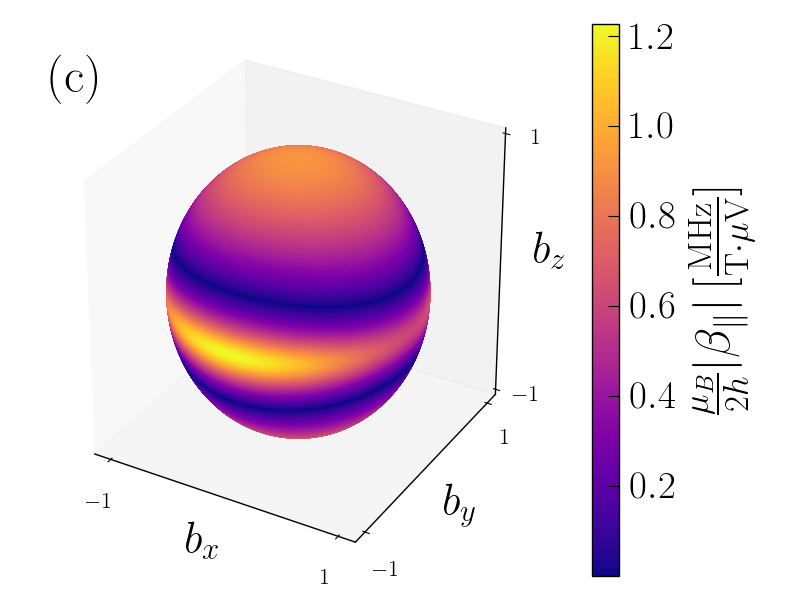}
\caption{Maps of (a) the effective $\gt$-factor and (b), (c) the perpendicular $\beta_{\perp}$ and parallel $|\beta_{\parallel}|$ susceptibilities as a function of the magnetic field orientation $\mathbf{b}=\mathbf{B}/|\mathbf{B}|$. We consider an idealized silicon device with hard-wall confinement along $y$ and $z$ ($L_z=15\unit{nm}$, $L_y=40\unit{nm}$), harmonic confinement along $x$ (harmonic length $\ell_x=11.5\unit{nm}$, as defined in~\cite{Michal2021}), and a homogeneous static electric field $\ef_y=0.12\unit{mV/nm}$ (we assume $\partial \ef_y/\partial V=-1/L_y$ \cite{suppmat}). The $\gt$-matrices are numerically computed from the solutions of the four bands Luttinger-Kohn Hamiltonian ~\cite{Venitucci2019,Michal2021,suppmat}. The longitudinal susceptibility $|\beta_{\parallel}|$ is maximum for specific orientations (dephasing hot spots) where the transverse susceptibility $\beta_{\perp}$ is zero, and vanishes along dephasing sweet lines. The transverse susceptibility $\beta_{\perp}$ is maximum (fastest Rabi oscillations) very close to such a sweet line.} 
\label{fig:gpbparaperp}
\end{figure*}

In the thin dot regime $L_z\ll L_{x,y}$, the spin wave-function has a dominant heavy-hole character~\cite{Katsaros2011,Ares2013} along the strong confinement axis taken as $z=[001]$. The behavior of the $\gt$-matrix of such heavy-hole devices has been extensively analyzed with different methods~\cite{Venitucci2018,Venitucci2019,Michal2021,martinez2022,LilesPRB2021} reproducing experimental observations~\cite{Crippa2018,Piot2022}. We provide details about our theoretical derivation of the $\gt$-matrix in the Supplemental Material (SM)~\cite{suppmat}. Given the freedom of choice on the magnetic axes and two-level basis set, the $\gt$-matrix can always be cast in diagonal form 
\begin{equation}\label{eq:diagg}
\gt(V)=\mbox{diag}\,[\,\gt_x\,,\,\gt_y\,,\,\gt_z\,]\,,
\end{equation}
where $\gt_z>\gt_{x,y}$ is the fingerprint of the dominant heavy-hole character of the wave-function, while the smaller $\gt_{x,y}$ arise from heavy/light hole mixing, mainly controlled by lateral confinement in the $(xy)$ plane. Ultimately, this $\gt$-matrix leads to a strongly anisotropic Larmor pseudovector $\boldsymbol{\omega}_L=\tfrac{\mu_B}{\hbar}\gt\cdot\mathbf{B}$~\cite{Abragam1970}. Figure~\ref{fig:gpbparaperp}a illustrates this anisotropy for a particular realization of the device of Fig.~\ref{fig:setup}, by showing the effective $\gt$-factor $\gt^\ast=|\gt\cdot\mathbf{b}|$ relating the Larmor angular frequency $|\boldsymbol\omega_L|=\tfrac{\mu_B}{\hbar}\gt^\ast B$ to the amplitude $B=|\mathbf{B}|$ of the magnetic field for a given orientation $\mathbf{b}=\mathbf{B}/B$.

We focus now on the coupling of the hole spin to photons confined in a microwave resonator with angular frequency $\omega_0$. These photons give rise to quantized voltage fluctuations $\delta V=V_{\rm zpf}(a+a^\dagger)$ on the confining gate, where $V_{\rm zpf}=\omega_0\sqrt{\hbar Z_r/\pi}$ is the zero point voltage fluctuation at the resonator edge, $Z_r$ is its characteristic impedance and the operator $a^\dagger$ creates a photon~\cite{devoretCircuitQEDHowStrong2007,Childress2004}. Such voltage fluctuations modulate the $\gt$-matrix, leading to spin-photon coupling. Without loss of generality, we separate the spin precession vector $\boldsymbol \sigma$ into components longitudinal ($\parallel$) and transverse ($\perp$) to the static Larmor vector $\boldsymbol{\omega}_L(\vg)$ \cite{Kato2003,Ithier2005}, where $\vg$ is the DC bias on the gate. Expanding Eq.~\eqref{eq:diagg} to the leading order in $V_{\rm zpf}$, the Hamiltonian for the spin-resonator dynamics reads
\begin{equation}\label{H_main}
\mathcal H=\frac{\hbar}{2}\omega_L\sigma_\parallel
+\hbar\omres a^\dagger a+\hbar\left[g_\parallel\sigma_\parallel+g_\perp\sigma_\perp\right](a+a^\dagger)\,,
\end{equation} 
where $\omega_L=|\boldsymbol{\omega}_L(\vg)|$. The spin-photon coupling thus involves longitudinal and transverse components, see Fig.~\ref{fig:setup}b. Defining the unit vector $\mathbf n$ along $\boldsymbol\omega_L(\vg)$, these couplings take the form 
\begin{equation}\label{eq:gparaperp}
 g_{\parallel/\perp}=\vzpf\frac{\mu_B B}{2\hbar}\beta_{\parallel/\perp}(\mathbf{b})\,,
\end{equation}
where we have introduced the parallel and transverse $\gt$-susceptibilities
\begin{align}\label{eq:betas}
\beta_{\parallel}&=\left(\gt'(\vg)\cdot\mathbf{b}\right)\cdot\mathbf{n}\,,&\beta_{\perp}&=\left|\left(\gt'(\vg)\cdot\mathbf{b}\right)\times\mathbf{n}\right|\,,
\end{align}
with the shorthand notation $\gt'(\vg)=\partial_V\gt(V)|_{V=\vg}$. These susceptibilities characterize the spin-photon couplings per unit of voltage and magnetic field.

The transverse ($\beta_\perp$) and longitudinal ($\beta_\parallel$) susceptibilities show strong dependence on the orientation of the magnetic field $\mathbf B$, as illustrated in Figs.~\ref{fig:gpbparaperp}b-c. 
In particular, $\beta_\perp$ vanishes where $\beta_\parallel$ is maximum (longitudinal sweet spot, along the device axes), while $|\beta_\parallel|$ vanishes next to the maxima of $\beta_\perp$ (transverse sweet spot). At such sweet spots, the hole spin is better protected against, respectively, relaxation ($\beta_\perp=0$) or dephasing ($\beta_\parallel=0$) due to electrical noise \footnote{\mf{When $\beta_\perp$ vanishes, the spin can not be manipulated by electric-dipole spin resonance with the top gate. It may still be manipulated, however, by other (e.g., side) gates \cite{Michal2021}.}}. This ``reciprocal sweetness'' between the transverse and longitudinal couplings is a key argument of the following discussion and has a robust geometrical origin. Indeed, $\beta_\parallel$ and $\beta_\perp$ are proportional to the projections of $\gt'\cdot\mathbf b$ onto vectors that are respectively parallel and perpendicular to $\boldsymbol\omega_L$. Therefore, whenever one of the couplings is zero because $\gt'\cdot\mathbf b$ and $\boldsymbol\omega_L$ are either parallel or perpendicular, the other is expected to be maximum. \mf{Reciprocal sweetness shall hence be ubiquitous in a large variety of electron and hole spin devices that can be described by the $\gt$-matrix formalism}. The role played by the variations of $|\gt'\cdot\mathbf b|$ in the reciprocal sweetness are discussed in the SM~\cite{suppmat}. 

\mf{A non-zero $g_\parallel$ here means that gate voltage fluctuations effectively modulate the Zeeman splitting $|\boldsymbol\omega_L|$ by reshaping the hole wave function. This mechanism is known as $\gt$-tensor modulation resonance ($\gt$-TMR) in the context of Rabi oscillations (transverse coupling). We point out, though, that additional contributions to $g_\parallel$ may arise from the displacement of the dot as a whole \cite{bosco2022fully}. They are negligible at low lateral electric field $\ef_y$ in the present setup but would be relevant if the hole was driven along the wire axis $x$ at large $\ef_y$ \cite{Michal2021}, allowing for significant Rashba-type SOC} \footnote{\mf{Nevertheless, the present $\gt$-matrix formalism catches the contribution of Rashba-type SOC to resonant transverse processes (Rabi oscillations and resonant protocols based on $g_\perp$) \cite{Venitucci2018}.}}.

\begin{figure}[b]
\centering
  \includegraphics[width=\columnwidth]{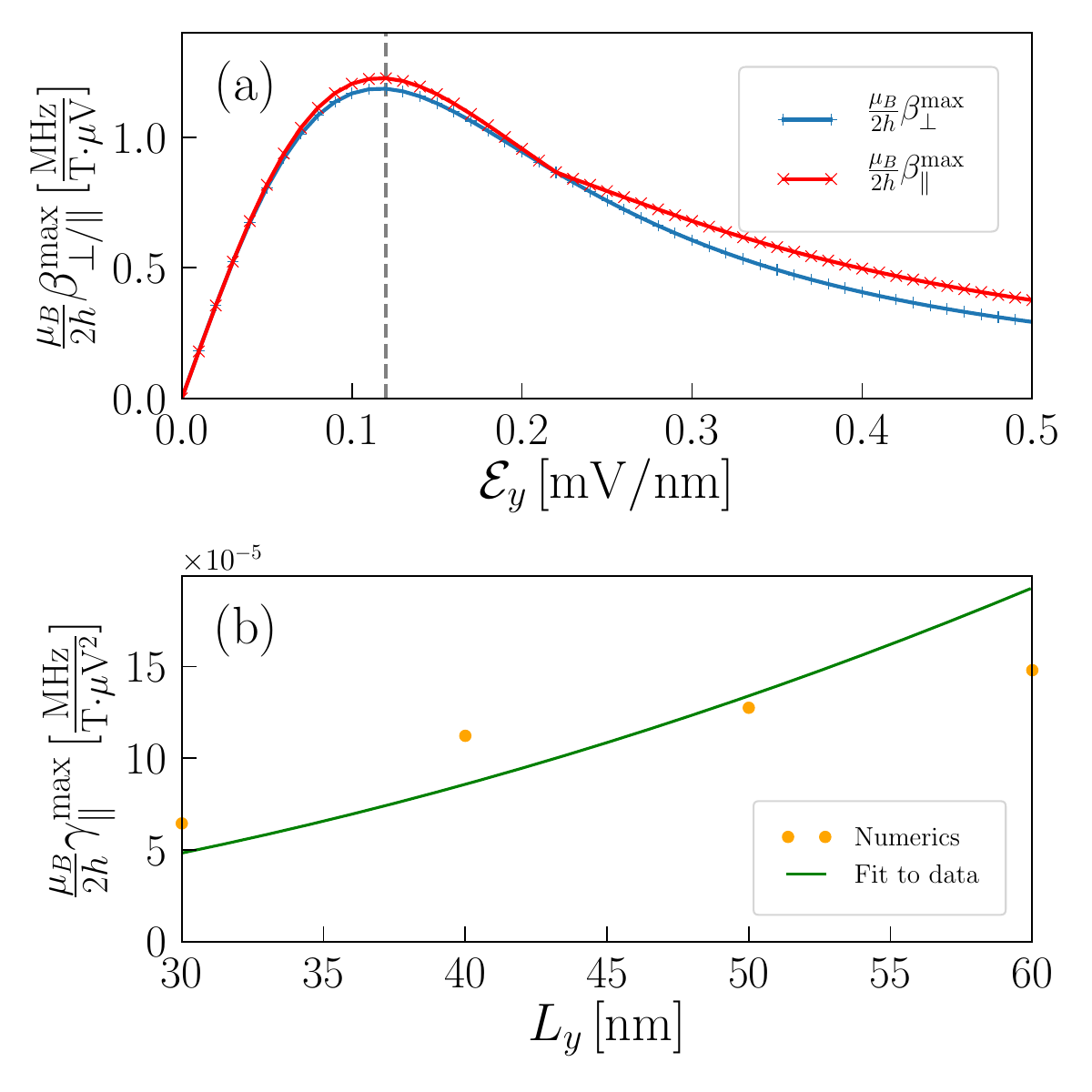}
\caption{(a) Maximum parallel and perpendicular susceptibilities $\beta_{\parallel}^{\max}$ (red) and $\beta_{\perp}^{\max}$ (blue) as a function of the static electric field $\ef_y$, for the same parameters as in Fig.~\ref{fig:gpbparaperp}. They are computed at the magnetic field orientations that maximize respectively $\beta_{\parallel}$ and $\beta_{\perp}$ for each $\ef_y$. The dashed line pinpoints the electric field $\ef_y^\ast$ that optimizes $\beta_{\parallel}^{\max}$ and $\beta_{\perp}^{\max}$. 
(b) Second-order longitudinal susceptibility $\gamma_\parallel$ at the optimal $\ef_y=0$ and $\mathbf{B}\parallel\mathbf{y}$, as a function of $L_y$. The dots are the numerical calculations with the parameters of Fig. \ref{fig:gpbparaperp}, and the line is an analytical fit $\gamma_{\parallel}^{\max}=0.14m_\parallel^h L_y^2e^2/(\pi^3\hbar^2\Delta)$.}
\label{fig:gpbmaxE}
\end{figure}

In the device of Fig.~\ref{fig:setup}, the deformation of the wave-function is controlled by the joint action of the lateral electric field $\ef_y$ and the anharmonic hard-wall confinement across the channel width. The interplay between anharmonicity and symmetries is illustrated in Fig.~\ref{fig:gpbmaxE}a, where we track the maximum $\beta_{\parallel/\perp}^{\max}$ as a function of the lateral electric field $\ef_y$. In the absence of electric field ($\ef_y=0$), both $\beta_\parallel$ and $\beta_\perp$ vanish due to the existence of an inversion symmetry in the confinement potential~\cite{Venitucci2018}. Once the electric field is turned on, the hole gets progressively squeezed onto the side facet of the channel, and the spin-photon couplings increase. Furthermore, $\beta_{\parallel}^{\max}$ and $\beta_{\perp}^{\max}$ are of the same order of magnitude whatever $\ef_y$. Ultimately, $\beta_{\parallel/\perp}$ tend to zero at large $\ef_y$ as the heavily squeezed hole hardly responds any more to fluctuations of the electric field. Both spin-photon couplings are optimal at a comparable electric field $\ef_y^\ast=0.12\unit{mV/nm}$ \footnote{\mf{$\gt$-TMR is, therefore, typically optimal at much lower electric fields than Rashba SOC \cite{Michal2021}. The depth of the dot and the lateral electric field may be adjusted independently with, e.g., a back gate, as shown in \cite{suppmat}.}}, with $\mathbf B$ lying in the $(xy)$ plane for $\beta_\parallel$, and at an angle $\delta\theta=\pm \arctan(\sqrt{|\gt_{y}/\gt_z|})$ from that plane for $\beta_\perp$~\cite{Venitucci2019,Michal2021}. We emphasize that $\beta_\perp$ is typically maximal at out-of-plane values of $B_z$ that remain compatible with the critical fields of resilient superconducting resonators~\cite{HighBresonatorSamkharadze, HighBresonatorYu}.

We confirm analytically that $\beta_{\parallel}^{\max}=\beta_{\perp}^{\max}$ in the thin dot regime $L_z\ll L_{x,y}$~\cite{suppmat}. The optimal susceptibilities $\beta_{\parallel/\perp}^\ast=Ce/\Delta$ are achieved at the electric field $e\ef_{y}^{\ast}\sim1.5\pi^3\hbar^2/(m_\parallel^h L_y^3)$, where $\Delta$ is the heavy-hole/light-hole band gap, $m_\parallel^h$ is the in-plane heavy-hole effective mass, and $C$ is a material-dependent constant of order unity ($C\approx 0.33$ for silicon and $C\approx 0.84$ for germanium).

The thin dot regime is, however, usually suboptimal as the gap $\Delta\propto L_z^{-2}$ may be large. In the more favorable conditions of Fig.~\ref{fig:gpbparaperp}, and considering a typical $\vzpf=20\unit{\mu V}$, the numerically computed couplings at $\ef_y^\ast\simeq0.12\unit{mV/nm}$ reach up to $g_{\parallel/\perp}/2\pi\sim 10\unit{MHz}$ at $B=0.5\unit{T}$ ($\omega_L/2\pi\approx 5\unit{GHz}$ for parallel and $\omega_L/2\pi\approx12\unit{GHz}$ for perpendicular coupling). We stress that these values for single holes confined in a single quantum dot are comparable to the transverse couplings reported in spin-cQED experiments with electrons in silicon double quantum dots with micromagnets~\cite{Samkharadze2018,Mi2018}. We show in the SM that these estimates are reproduced by numerical simulations of more realistic devices with inhomogeneous lateral and vertical electric fields~\cite{suppmat}.

%%%%%%%%%%%%%%%%%%%%%%%%%
% APPLICATIONS
%%%%%%%%%%%%%%%%%%%%%%%%%

The strong dependence of $g_{\parallel/\perp}$ on the electric field $\ef_y$ can be used to turn on and off the coupling between single hole spins and resonators. Moreover, owing to the reciprocal sweetness between $\beta_{\parallel}$ and $\beta_{\perp}$, the nature of the spin-photon coupling can be switched from purely transverse to purely longitudinal by a rotation of the magnetic field. This provides a playground for the exploration of the physics and applications of spin-photon interactions in arbitrary regimes\mf{, and allows the optimization of a specific coupling for the operations targeted in an experiment.} 

As an example, the sweet lines for transverse coupling ($\beta_\parallel=0$) enable spin-microwave photon quantum state transfer \cite{Beaudoin2016} with fidelities reaching $\sim 97\%$ for realistic noise, typically limited by the photon losses in the resonator (see SM~\cite{suppmat}). Single holes have already demonstrated coherence times of the order of tens of $\mu$s on such sweet lines \cite{Piot2022}. In principle, it is also possible to leverage the longitudinal coupling at finite $\ef_y$ to mediate CZ gates or ZZ interactions between distant qubits~\cite{roos2008ion,Jin2012,Harvey2018,Schuetz2017,Schuetz2019,suppmat}. Nevertheless, regimes where $\beta_\parallel$ is large are hot spots for two-level dephasing induced by charge noise~\cite{Connors2019,Chanrion2020,Piot2022}, which may compromise the experimental demonstration of coherent spin manipulation. 

This difficulty can be overcome by exploiting the second-order longitudinal coupling proportional to the {\it curvature} $\gt''(\vg)$ of the $\gt$-matrix~\cite{Didier2015,roos2008ion, Kerman2013, Billangeon2015, Richer2016, Royer2017, Schuetz2017, Schuetz2019, Harvey2018, Bottcher2021, Ruskov2019, Ruskov2021}. Curvature coupling indeed prevails at $\ef_y=0$, where $\beta_\parallel=\beta_\perp=0$, so that the qubit is insensitive (to first-order) to both charge noise dephasing and relaxation.

In this regime, if the resonator is driven at its resonant frequency $V\rightarrow V(t)=\vg+V_{\rm d}\cos(\omega_0 t)$, the Hamiltonian~\eqref{H_main} reduces to $\mathcal H=\frac{\hbar}{2}\tilde g_\parallel\sigma_\parallel(a+a^\dagger)$ in the frame rotating at angular frequency $\omega_0$~\cite{Didier2015}. There the parametrically strong longitudinal coupling $\tilde g_\parallel$ is proportional to the curvature of the $\gt$-matrix ~\footnote{We stress that Eq.~\eqref{eq:curvature1} assumes $\beta_{\parallel/\perp}=0$. A more general expression valid for regimes where $\beta_{\parallel/\perp}\neq0$ is given in the SM~\cite{suppmat}.}:
\begin{align}\label{eq:curvature1}
 \tilde{g}_{\parallel}&=\vzpf V_{\rm d}\frac{\mu_BB}{2\hbar}\gamma_{\parallel}\,,&\gamma_{\parallel}&=\left(\gt''(\vg)\cdot\mathbf{b}\right)\cdot\mathbf{n}\,,
\end{align}
where we use the shorthand notation $\gt''(\vg)=\partial^2_V\gt(V)|_{V=\vg}$. Notice that there is a concomitant transverse curvature parameter $\gamma_{\perp}$ in total analogy with Eq.~\eqref{eq:betas}. Nevertheless, reciprocal sweetness applies to $\gamma_{\parallel/\perp}$ for the same geometrical reasons as for $\beta_{\parallel/\perp}$. Namely, the orientation of the magnetic field $\mathbf B$ can be always chosen such that $\gamma_\perp=0$ while $\gamma_\parallel$ is maximum, or vice-versa. The dependence of $\gamma_{\parallel/\perp}$ on the magnetic field orientation is actually similar to that shown in Fig.~\ref{fig:gpbparaperp} for the linear susceptibilities $\beta_{\parallel/\perp}$~\cite{suppmat}. 

The curvature coupling can be readily tuned (and switched on/off) with the drive amplitude $V_d$. We also highlight that $\gamma_{\parallel}$ strongly depends on the lateral confinement length $L_y$, and tends to increase when widening the channel as shown in Fig.~\ref{fig:gpbmaxE}b. This trend can indeed be understood from Fig.~\ref{fig:gpbmaxE}a, which suggests that $\gamma_\parallel^{\max}\sim\beta_{\parallel}^{\rm max}/(L_y \ef_y^*)\sim C m_\parallel^h L_y^2e^2/(\pi^3\hbar^2\Delta)$ in the thin dot regime. This estimate is however expected to be less accurate when $L_z$ gets comparable to $L_y$, as is the case in Fig.~\ref{fig:gpbmaxE}b, where we actually fit the numerical data with $C=0.14$. \mf{Moreover, the curvature coupling tends to saturate at large $L_y\gg\ell_x/\pi$.} Practically, the quantum dot size will ultimately be limited by disorder-induced localization~\cite{martinez2022}, and/or by the larger magnetic fields needed to compensate for the decrease of $\gt_y$.

For the device parameters of Figs.~\ref{fig:gpbparaperp} and~\ref{fig:gpbmaxE}, 
assuming $V_{\rm zpf}=20\unit{\mu V}$, a magnetic field $B=2\unit{T}$ along $y$ ($\omega_L/2\pi\sim 6\unit{GHz}$), and a voltage drive $V_{\rm d}=2\unit{mV}$, the longitudinal coupling is of the order of $\tilde{g}_{\parallel}/2\pi\sim 
10\unit{MHz}$. Such couplings would enable different operations, such as genuine quantum non-demolition measurements~\cite{Didier2015}, CZ gates and, more generally, non-resonant ZZ interactions between distant spins connected to the same resonator~\cite{roos2008ion, Jin2012, Royer2017, Harvey2018, Schuetz2019}. We address the fidelity of a CZ gate leveraging the curvature coupling in the SM~\cite{suppmat}, and find that the error may be brought down to the percent range thanks to the high coherence at $\ef_y=0$.

In conclusion, we have shown that various aspects of spin circuit-QED can be efficiently addressed and exploited in minimal single hole semiconductor devices. The all-electrical modulation of the hole $\gt$-matrix allows to tune both transverse and longitudinal spin-photon couplings, whose relative strengths can be controlled by the orientation of the magnetic field thanks to their reciprocal sweetness. In an experiment the couplings can be electrically switched on and off, either by using the gate voltage dependence of the $\beta_{\parallel/\perp}$ susceptibilities (see Fig.~\ref{fig:gpbmaxE}), or by switching-off the AC modulation of the gate voltage for parametric longitudinal coupling. The transverse coupling can be harnessed to transfer the spin state to the microwave photon state, with high fidelity when operating on the sweet lines where dephasing due to electric noise is reduced. On the other hand the longitudinal coupling can be used for fast quantum non-demolition readout and robust multi-spin interactions~\cite{Billangeon2015, Didier2015}. 

An interesting perspective would be to tune the relative strength of the transverse and longitudinal couplings with a fully electrical knob instead of the magnetic field orientation. This would enable rapid switching between both couplings within single experiments, but calls for the ability to rotate electrically the spheres in Fig.~\ref{fig:gpbparaperp}b-c around the $x$ or $y$ axis. This operation is challenging in simple, few-gate geometries and we leave this question open for further investigations.

While we have mainly considered single holes in silicon nanowire quantum dots, the principles discussed in this work also apply to other hole spin qubit systems, such as germanium nanowires and heterostructures \cite{Mutter2020, Bosco2021}, or dopant-based hole qubits \citep{salfi2016charge, abadillo2018entanglement}. Further improvements in the quality of the microwave resonators and the use of low-loss substrates will lead to enhancements of the fidelities of multi-spin operations and allow progress towards a highly connected spin-qubit architecture for quantum computation and quantum simulation. 

{\it Acknowledgements  -- }We thank  Olivier Buisson, Silvano de Franceschi, \'Etienne Dumur and Benoît Vermersch for useful comments and discussions. This work was supported by the French National Research Agency (ANR) through the MAQSi project, and by the European Union’s Horizon 2020 research and innovation program under grant agreements No. 810504 (ERC project QuCube) and No. 759388 (ERC project LONGSPIN).

\bibliographystyle{apsrev4-2}    
\bibliography{biblio}

\newpage

\onecolumngrid

\section*{Supplemental Material}

Here we give details on the hole $\gt$-factors in the quasi-two-dimensional regime (Section~\ref{sec:thin_channel}), on the fidelity of the spin/microwave-photon quantum state transfer protocol in the presence of spin/photon decay and quasistatic charge noise (Section~\ref{sec:spintransfer}), and on the fidelity of the two-qubit controlled-Z gate based on longitudinal couplings (Section~\ref{sec:longitudinalfidel}). Finally, we further support the proposal with numerical simulations on a more realistic setup (Section~\ref{sec:realistic}). 

\subsection{Holes in a thin channel}\label{sec:thin_channel}

The hole $\gt$-factors in the quasi-two-dimensional (thin dot) regime were analytically calculated in~\cite{Michal2021} with the four bands Luttinger-Kohn Hamiltonian. There, the analytical expressions were benchmarked against numerical methods developed in \cite{Venitucci2018,Venitucci2019}. For holes with mostly heavy character, only the diagonal elements of the $\gt$-matrix are non-zero and read:
\begin{subequations}
\begin{align}
\gt_{x}&=\frac{6\gamma_3\hbar^2}{m_0\Delta}\left(\kthree\langle k_y^2\rangle-\ktwo\langle k_x^2\rangle\right)\,,\\
\gt_{y}&=\frac{6\gamma_3\hbar^2}{m_0\Delta}\left(\ktwo\langle k_y^2\rangle-\kthree\langle k_x^2\rangle\right)\,,\\
\gt_{z}&=-6\kappa+2\gamma_{h}+\delta\mathsf{g}_z\,.
\end{align}
\end{subequations}
Here $m_0$ is the bare electron mass, $\gamma_1$, $\gamma_2$, $\gamma_3$ are the Luttinger parameters characterizing the hole masses, and $\kappa$ is the valence band Zeeman parameter. $\Delta=2\pi^2\gamma_2\hbar^2/m_0L_z^2$ is the energy splitting between the heavy-hole and light-hole 2D subbands, and $\ktwo=\kappa-2\gamma_{2}\eta_h$, $\kthree=\kappa-2\gamma_{3}\eta_h$, where $\gamma_h$ and $\eta_h$ are dimensionless parameters defined in Refs.~\cite{Ares2013PRL,Michal2021} ($\gamma_h\approx 1.16$ and $\eta_h\approx 0.08$ in thin silicon films with hard wall boundary conditions). The expectations values of $k_x=-i\partial_x$ and $k_y=-i\partial_y$ are calculated for the ground-state heavy-hole envelope. The term $\delta\mathsf{g}_z$ collects corrections of order $\hbar^2\langle k_x^2\rangle/m_0\Delta$ and $\hbar^2\langle k_y^2\rangle/m_0\Delta$ \cite{Katsaros2011}. In this regime $\gt_x,\gt_y\ll\gt_z$ (see Figure 2 of the Main Text).

In this work, we assume (as in Ref. \cite{Michal2021}) hard-wall confinement in a rectangular nanowire with sides $L_y$ and $L_z$, and harmonic confinement along $x$ with characteristic length $\ell_x\propto L_x$ ($\ell_x\approx 11.5$ nm for gate length $L_x\approx 35$ nm \cite{Venitucci2018,Michal2021}). Following Ref. \cite{Michal2021}, we can then optimize the transverse and parallel susceptibilities with respect to the magnetic field orientation, and find $\beta_{\perp/\parallel}^{\max}=\max(\gt'_x(\vg),\gt'_y(\vg))$. $\beta_{\parallel}$ is actually maximum for a magnetic field $\mathbf B$ in the $(xy)$ plane, and $\beta_{\perp}$ for a magnetic field $\mathbf B$ at an angle $\delta\theta=\pm \arctan(\sqrt{|\gt_{y}/\gt_z|})$ from that plane. Further optimization with respect to the electric field $\ef_y$ yields:
\begin{equation}
\beta_{\parallel/\perp}^{\ast}\approx\frac{1.2\gamma_3\max(|\ktwo|,|\kthree|)e}{(\gamma_1+\gamma_2-\gamma_h)\Delta}\equiv\frac{Ce}{\Delta}\,,
\label{eq:betastar}
\end{equation}
at $e\ef_{y}\sim3\pi^3\hbar^2/(2m_\parallel^h L_y^3)$, where $m_\parallel^h=m_0/(\gamma_1+\gamma_2-\gamma_h)$ is the in-plane heavy-hole mass and $e>0$ is the elementary charge (see the corresponding equation in the Main Text). We have assumed here that zero point gate voltage fluctuations $\delta V=V_{\rm zpf}$ translate into zero point electric field fluctuations $\delta \ef_y=-V_{\rm zpf}/L_y$. The relevance of this assumption is confirmed by the simulations of section \ref{sec:realistic}. Eq.~\eqref{eq:betastar} is valid in the thin dot regime $\Delta\gg\hbar^2\gamma_3/(m_0\ell_{\parallel}^2)$, where $\ell_{\parallel}$ is the minimal in-plane confinement length, which may ultimately be limited by disorder \cite{martinez2022}. Beyond the thin dot regime, the Luttinger-Kohn equations can be solved numerically with Fourier-series expansions, as done in \cite{Venitucci2019,Michal2021} and in Figs. 2 and 3 of the Main Text.

\begin{figure}[h]
\centering
\includegraphics[width=0.4\textwidth]{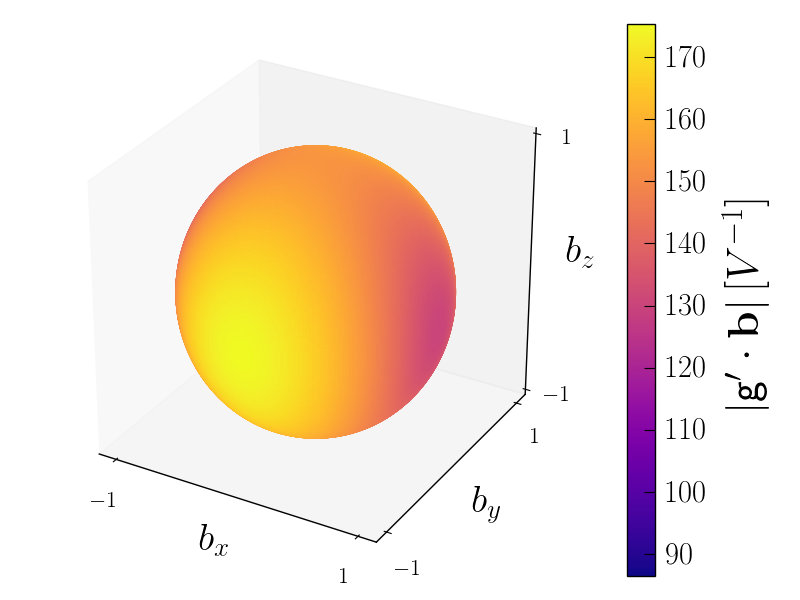}
\caption{Map of the norm of the Larmor vector derivative as a function of magnetic field orientation, for the same parameters as in Fig.~2 of the Main Text.}
\label{gpb_norm}
\end{figure}

As discussed in the Main Text, one of the susceptibilities $\beta_\parallel$ or $\beta_\perp$ is expected to be maximum if the other vanishes because $\gt'\cdot\mathbf b$ and $\boldsymbol\omega_L$ are either parallel ($\beta_\perp=0$) or perpendicular ($\beta_\parallel=0$). In practice however, the actual maxima of $|\beta_\parallel|$ or $\beta_\perp$ may be shifted from the zeros of the other because $|\gt'\cdot\mathbf b|$ also depends on $\mathbf b$ [see Eq.~(5) in the Main text]. In the present case, $|\gt'\cdot\mathbf b|$ is only weakly dependent on the orientation of the magnetic field, as shown in Fig.~\ref{gpb_norm}. The maxima of $|\beta_\parallel|$ therefore coincide with the zeros of $\beta_\perp$, because they lie on the high symmetry device axes, while the maxima of $\beta_\perp$ are very slightly moved away from the sweet lines of $\beta_\parallel$. In general, one may still expect a maximum of one coupling in the close vicinity of a zero of the other, unless the variations of $|\gt'\cdot\mathbf b|$ are really strong around that point.

\subsection{Fidelity for the resonant spin/microwave-photon state transfer in the presence spin/photon decay and quasistatic charge noise}\label{sec:spintransfer}

\begin{figure*}[b]
\centering
  \includegraphics[width=0.45\textwidth]{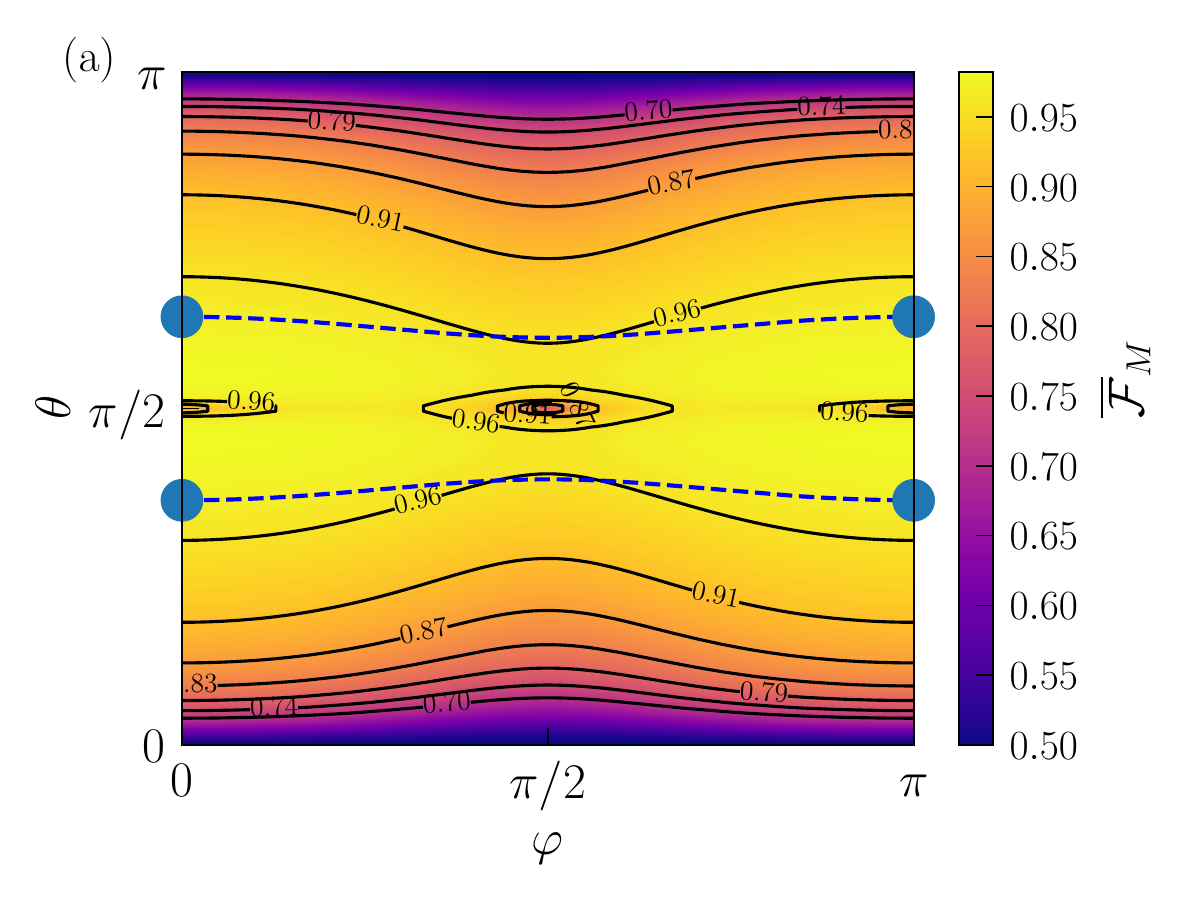}
\includegraphics[width=0.45\textwidth]{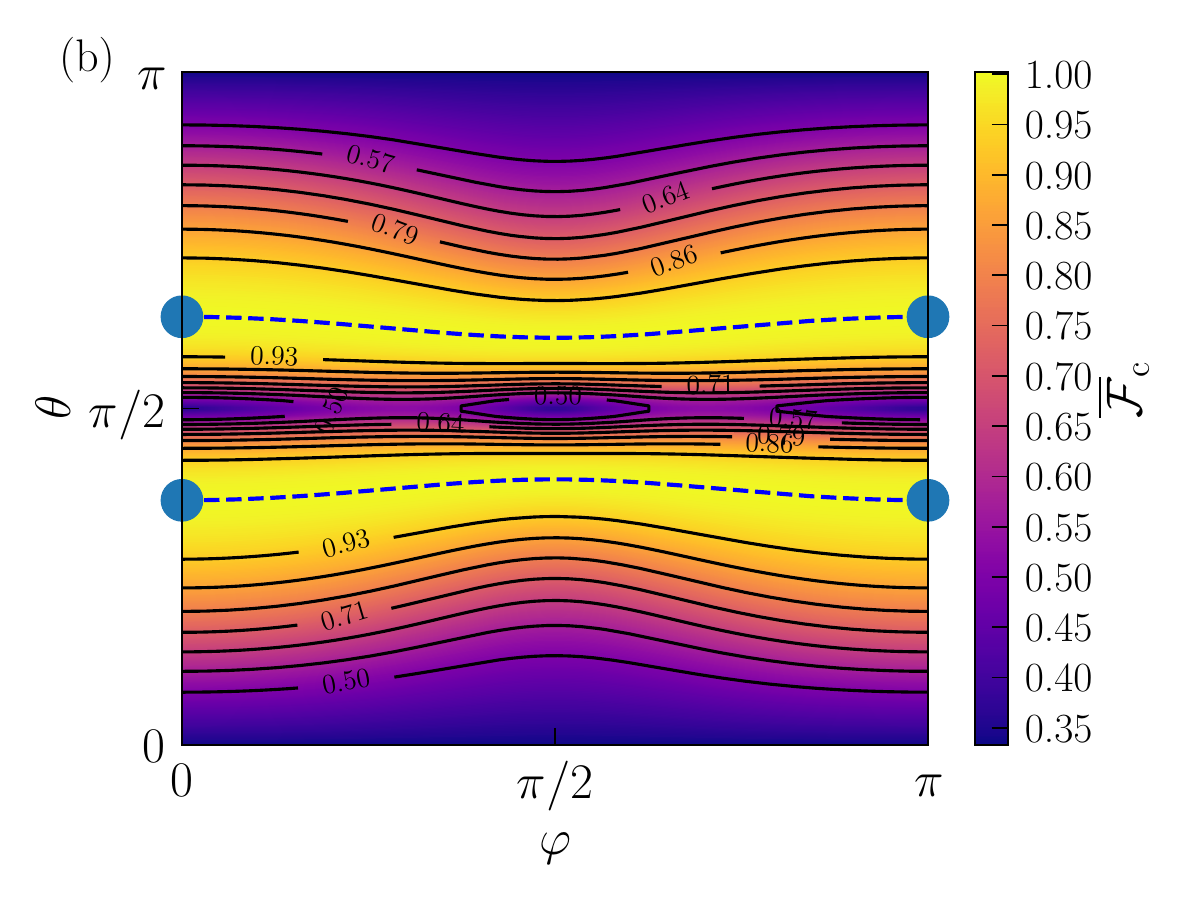}
\caption{Fidelity of the spin-microwave photon quantum state transfer protocol as a function of the orientation of the magnetic field, using the spin-photon couplings of Fig. 2 of the Main Text. We use the same convention for the magnetic field angles $\theta$ and $\varphi$ as in Refs.~\citenum{Venitucci2018,Venitucci2019,Michal2021}: $\theta$ is the angle between the $z$ axis and the magnetic field, and $\varphi$ is the angle between its projection in the $(xy)$ plane and the $y$ axis.
(a) Fidelity $\mathcal{\overline{F}}_M$ limited by the decay of the microwave photon as given by Eq.~(\ref{eq:Master}) with parameters $\vzpf=20\unit{\mu V}$ and $\kr/2\pi=0.5\unit{MHz}\gg \gamma_\downarrow/2\pi$, at constant Larmor frequency $\omega_L/2\pi=\omega_0/2\pi=5\unit{GHz}$. (b) Fidelity $\mathcal{\overline{F}}_c$ limited by quasistatic charge noise as given by Eq.~(\ref{Fcn}) with $\sigma_V=20\unit{\mu V}$~\cite{Connors2019,Chanrion2020}. The dashed blue lines are the sweet lines where the first-order longitudinal susceptibility $\beta_\parallel$ cancels and $\mathcal{\overline{F}}_c\approx 1$. The blue dots are the points on these lines that maximize $\mathcal{\overline{F}}_M\approx97\%$.}
\label{fig:fidelities}
\end{figure*}

As an example of a protocol realizable with the setup presented here, we discuss the spin-microwave photon quantum state transfer \cite{Beaudoin2016} and compute the fidelities achievable for realistic noise (Markovian and quasistatic). This protocol makes use of the transverse coupling to map the quantum state $|\psi_{\rm in}\rangle=\cos(\zeta/2)|g,0\rangle+e^{i\xi}\sin(\zeta/2)|e,0\rangle$ to  
$|\psi_{\rm tar}\rangle=\cos(\zeta/2)|g,0\rangle-ie^{i\xi}\sin(\zeta/2)|g,1\rangle$ by performing a resonant $\pi$ rotation between the states $|e,0\rangle$ and $|g,1\rangle$ over time $t_g=\pi/2g_\perp$ [here $|g,n\rangle$ and $|e,n\rangle$ are respectively the ground ($g$) and excited ($e$) qubit states with $n$ photons in the resonator]. For that purpose, the resonant transverse coupling $g_\perp$ between the hole spin and the microwave photon can be electrically switched on and off thanks to the voltage dependence of the hole $\gt$-factors (see Fig.~3 of the Main Text) \cite{Venitucci2018, Michal2021}. Let us evaluate the fidelity of such an operation in the presence of noise. We first consider the effect of damping due to photon decay and spin relaxation. The fidelity is then defined as $\overline{\fid}_M=\overline{\langle \psi_{\rm tar}|\rho(t_g)|\psi_{\rm tar}\rangle}$, where $\rho(t_g)$ is the density matrix at time $t_g$, and the overline denotes an uniform average over the Bloch sphere parametrized by the angles $\zeta$ and $\xi$. The master equation for the density matrix of the coupled spin qubit/microwave photon reads in the presence of photon decay and qubit relaxation:
\begin{equation} \label{eq:Master}
\dot{\rho}(t) = -\frac{i}{\hbar}[\mathcal{H}',\rho(t)] + \kr\mathcal{D}[a]\rho(t) + \gamma_{\downarrow} \mathcal{D}[\sigma_-]\rho(t)\,, 
\end{equation}
where $\kr$ is the photon decay rate, $\gamma_{\downarrow}$ is the relaxation rate of the spin qubit, $\mathcal{D}[O]\rho=O \rho O^{\dagger}-\frac{1}{2}(O^{\dagger} O \rho + \rho O^{\dagger} O)$ is the damping superoperator, and $\mathcal{H}'$ is the Hamiltonian that includes perturbative corrections such as the qubit Lamb shift. Equation~\eqref{eq:Master} is valid when $k_B T\ll \hbar\omres$, $k_B$ being the Boltzmann constant and $T$ the device temperature. This equation can be solved exactly, which yields the average fidelity $\overline{\fid}_M=\tfrac{1}{2}+\tfrac{1}{3}e^{-a/2}+\tfrac{1}{6}e^{-a}$, with $a=\pi(\kr+\gamma_\downarrow)/4g_\perp$. Hence, for $2g_\perp\ll\omres$ and to first order in $(\kr+\gamma_\downarrow)t_g$, the average error evaluates to
$1-\overline{\fid}_{M}=\pi(\kr+\gamma_\downarrow)/12 g_\perp$. Here we assume that the fidelity is realistically limited by the microwave photon decay \cite{Hendrickx2020_single_hole}, which for $\kr/2\pi=0.5\unit{MHz}$ (resonator quality factor $Q_r=\omres/\kappa_r=10^4$ with $\omega_0/2\pi=5\unit{GHz}$), results in a minimal process error of a few percents (see Fig. \ref{fig:fidelities}a). 

We furthermore analyze the error in the spin-photon transfer process due to quasistatic charge noise on the gate. The process fidelity due to such electric noise is defined as
\begin{equation}
\overline{\fid}_{c}=\overline{\langle \psi_{\rm tar}|{\rm E}[U_g(V_G)|\psi_{\rm in}\rangle\langle\psi_{\rm in}|U_g^\dagger(V_G)]|\psi_{\rm tar}\rangle}\,,
\end{equation}
where $U_g(V_G)$ is the evolution operator at gate voltage $V_G$ and $\av{.}$ is the ensemble average over the random gate voltage fluctuations $\delta V_G=V_G-\vg$ that we assume Gaussian. The unitary evolution of the state $|\psi_{\rm in}\rangle=\cos(\zeta/2)|g,0\rangle+e^{i\xi}\sin(\zeta/2)|e,0\rangle$ under Hamiltonian (3) of the Main Text close to the resonant condition $|\delta\omega_L|=|\omega_L(V_G)-\omres|\ll\omres$ yields $U_g(V_G)|\psi_{\rm in}\rangle=\cos(\zeta/2)|g,0\rangle+\sin(\zeta/2)e^{i\xi}\left([\cos(\omega_R t_g/2)-i\cos\alpha\sin(\omega_R t_g/2)]|e,0\rangle-i\sin\alpha\sin(\omega_R t_g/2)|g,1\rangle\right)$, with $\cos\alpha=\delta\omega_L/\omega_R$, $\sin\alpha=2g_\perp/\omega_R$, and $\omega_R=\sqrt{4g_\perp(V_G)^2+\delta\omega_L(V_G)^2}$.
This leads to the fidelity of the quantum state transfer averaged over the initial state: 
\begin{equation}\label{Fcn}
\overline{\fid}_{c}=\frac{1}{3}\left(1+\av{\sin\alpha\sin(\omega_R t_g/2)}+\av{\sin^2\alpha\sin^2(\omega_R t_g/2)}\right)\,.
\end{equation}
For small deviation from resonance $|\delta\omega_L|\ll 2g_\perp(\vg)$ and $|\delta g_\perp|\ll g_\perp(\vg)$, the angular frequency of state rotation approximates as  
$
\omega_R\approx 2g_\perp(\vg) + 2\delta g_\perp + \delta\omega_L^2/4g_\perp(\vg),
$
and at leading order in the noise
$1-\overline{\fid}_c=(\av{\delta\omega_L^2}+\pi^2\av{\delta g_\perp^2})/8g_\perp^2$. Close to the optimal $g_\perp$ ($\ef_y\simeq\ef_y^\ast$), the second term of the latter expression is much smaller than the first one, so that 
\begin{equation}\label{fidc}
1-\overline{\fid}_c=\frac{1}{2}\left(\frac{\sigma_V \beta_{\parallel}}{\vzpf\beta_{\perp}}\right)^2\,,
\end{equation}
with $\sigma_V^2={\rm E}[\delta V_G^2]$ the variance of $\delta V_G$. A sweet condition occurs when the longitudinal coupling vanishes ($\beta_\parallel=0$) and the error cancels at this leading order in $\sigma_V$. Figure~2 of the Main Text shows the presence of sweet lines in the vicinity of the magnetic field orientations where the transverse coupling is maximal. As discussed previously, such sweet regions are expected to occur in a variety of systems and geometries \cite{Bosco2021,Piot2022}. In order to assess the robustness of the sweet lines, we compute the fidelity, Eq.~(\ref{fidc}), up to second order in gate voltage fluctuations. Then $\delta\omega_L=\omega_L(V_G)-\omres=\delta\omega_\parallel+\delta\omega_\perp^2/2\omres$, where the transverse noise is first-order with respect to the fluctuations, $\om_{\perp}=\frac{\mu_B B}{\hbar}\beta_{\perp}\delta V_G$, while the longitudinal noise (as well as $\delta g_\perp$) is expanded up to the second order,
$\om_{\parallel}=\om_{\parallel}^{(1)}+\om_{\parallel}^{(2)}=\frac{\mu_B B}{\hbar}\left[\beta_\parallel\delta V_G+
\frac{1}{2}\gamma_\parallel\delta V_G^2\right]$. Figure~\ref{fig:fidelities}b displays the spin-photon quantum state transfer fidelities numerically calculated this way. It reveals that the sweet lines are remarkably robust to higher order charge noise so that the process fidelity is overall limited by the resonator microwave photon decay.

\subsection{Fidelity of the CZ gates based on longitudinal first-order and second-order parametric couplings}\label{sec:longitudinalfidel}

\begin{figure}[h]
\centering
  \includegraphics[width=0.32\textwidth]{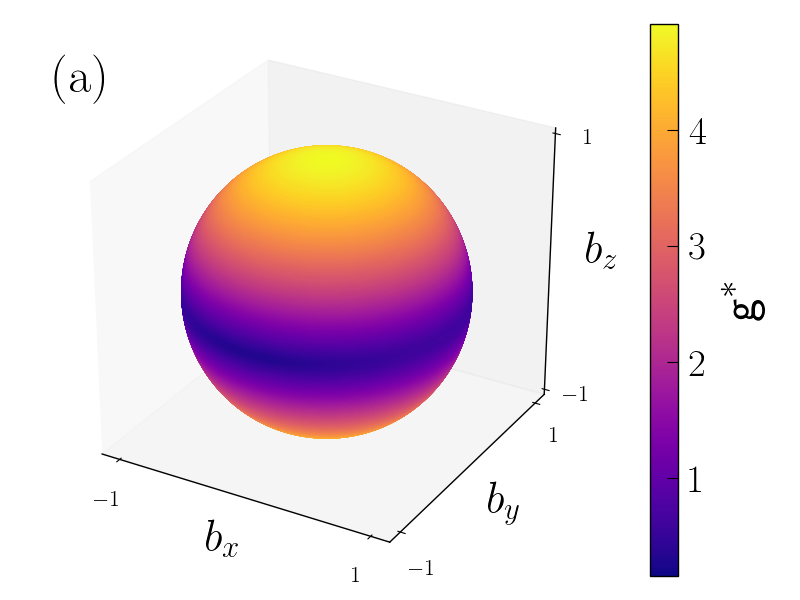}
  \includegraphics[width=0.32\textwidth]{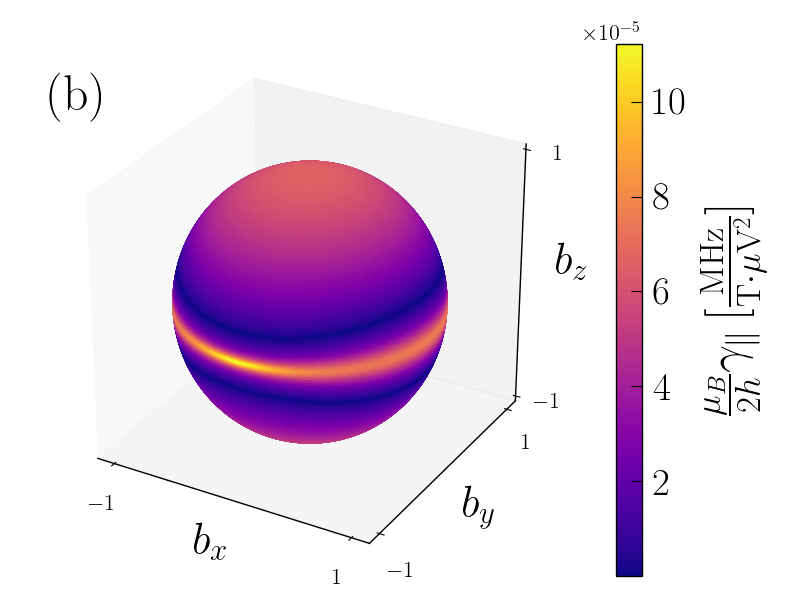}
  \includegraphics[width=0.32\textwidth]{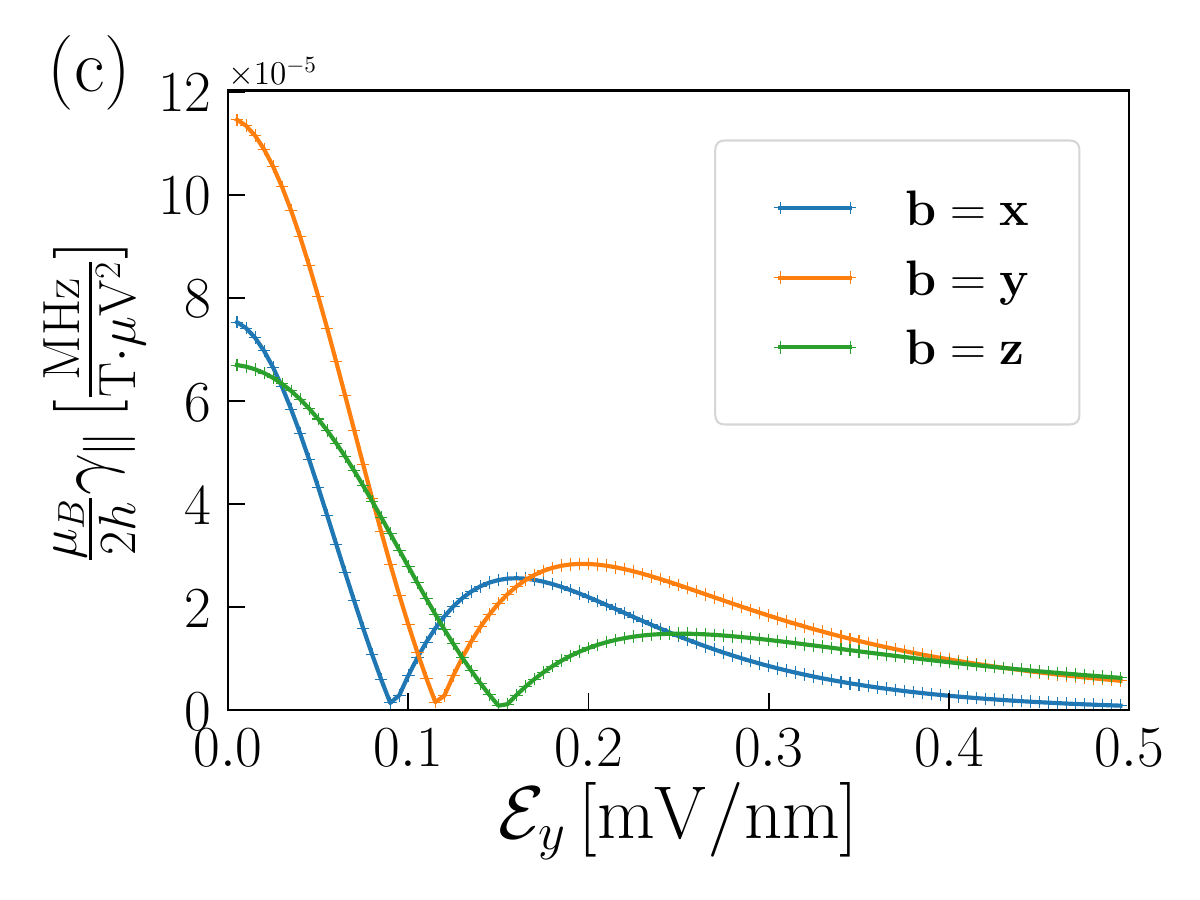}
\caption{(a) Effective $\gt$-factor and (b) curvature parameter $\tfrac{\mu_B}{2\hbar}\gamma_\parallel$ as functions of the orientation of the magnetic field at $\ef_y=0$. The parameters are the same as in Fig. 2 of the Main Text. (c) Curvature parameter $\tfrac{\mu_B}{2\hbar}\gamma_\parallel$ as a function of the electric field $\ef_y$ for different orientations of the magnetic field. $\gamma_\parallel$ is optimal for $\mathbf B$ along $y$ and for $\ef_y=0$ (where the first-order spin-photon couplings $g_{\parallel/\perp}$ are zero).}
\label{fig:longitudinal}
\end{figure}

So far, we have dealt with the coherent interaction between a microwave photon and a single-hole qubit using the most common transverse coupling. However, at particular magnetic field orientations, the single-hole qubit also exhibits pure longitudinal spin-photon coupling \cite{roos2008ion, Jin2012, Kerman2013, Billangeon2015, Richer2016, Royer2017, Schuetz2017, Schuetz2019, Harvey2018, Bottcher2021, Ruskov2019, Ruskov2021}, either at first order (Fig. 2 of the Main Text) or at second order (curvature or dispersive coupling, see Fig. \ref{fig:longitudinal}) with respect to the gate voltage. We comment on the formal correspondence between the two cases after Eq.~\eqref{fidcz} below. 

The longitudinal coupling provides a mechanism for efficient readout \cite{Didier2015} and for two-qubit gates based on geometric phases \cite{roos2008ion, Harvey2018, Schuetz2019}. The CZ gate mediated by longitudinal coupling for example has the important advantage to allow for efficient operation even when the resonator state is thermal \cite{Schuetz2017, Schuetz2019}, so that the latter does not need to be cooled down to its ground-state. Moreover, the static Larmor frequencies of the qubits can be different, since the longitudinal interaction can be harnessed non-resonantly. We focus here on the curvature coupling because of its tunability (the interaction is controlled by a drive) \cite{Harvey2018, Schuetz2019}, of the enhanced coherence (the curvature is optimal where the first-order susceptibilities $\beta_{\parallel/\perp}$ are zero), and of its relation to readout \cite{Didier2015}.  

We address below the fidelity of the CZ gate. For that purpose, we consider an ensemble of spins located at the anti-nodes of some mode $n$ of the resonator with angular frequency $\omega_n=(n+1)\omega_0$ (not necessarily the fundamental mode $n=0$). The longitudinal coupling between these spins can be tuned and possibly enhanced by applying a time-dependent voltage $V_d(t)=V_d\cos(\omega_d t)$ on the gates, with angular frequency $\omega_d$ close to $\omega_n$ \cite{Harvey2018, Schuetz2019}. The modulated longitudinal coupling of spin $i$, $g_{\parallel n}^{(i)}(t)=V_n\tfrac{\mu_B B}{2\hbar}\beta_{\parallel}^{(i)}+\tilde{g}_{\parallel n}^{(i)}\cos(\omega_d t)$, acquires a time-dependent component proportional to the curvature of the Larmor frequency:
\begin{equation}
 \tilde{g}_{\parallel n}^{(i)}=\frac{1}{2}\frac{\partial^2\omega_{L}^{(i)}}{\partial \vg^2}V_n V_{\rm d}\,,
\end{equation}
where 
\begin{equation}
\frac{\partial^2\omega_{L}^{(i)}}{\partial \vg^2}=\frac{\mu_B B}{\hbar}\gamma_\parallel
\end{equation}
and
\begin{equation}\label{eq:curvature2}
\gamma_\parallel=\left(\gt^{(i)''}(\vg)\cdot\mathbf{b}\right)\cdot\mathbf{n}+\frac{\left(\beta_{\perp}^{(i)}\right)^2}{\left|\gt^{(i)}\cdot\mathbf{b}\right|}\,,
\end{equation}
with $V_n=\sqrt{n+1}\vzpf$ the voltage amplitude of mode $n$ at a resonator anti-node \cite{Childress2004}.

It has been shown that for two spins coupled to a common resonator a polaron transformation \cite{roos2008ion, Jin2012, Royer2017, Harvey2018, Schuetz2019} yields the effective, non-resonant two-spin interaction $H_{\rm 12}=\hbar J_{12}\sigma_{\parallel}^{(1)}\sigma_{\parallel}^{(2)}$, where $J_{12}=-\tilde{g}_{\parallel n}^{(1)}\tilde{g}_{\parallel n}^{(2)}/2\Delta_n$ and $\Delta_n=\omega_n-\omega_d$. This interaction is realized at the discrete times $t_p=2\pi p/\Delta_n$ ($p=1,2,\dots$) at which the resonator decouples from the spins dynamics (assuming here $\Delta_n\,,\tilde{g}_{\parallel n}^{(i)}\ll\omega_d$). On the other hand, the two spins are maximally entangled at the two-qubit CZ gate time $t_{\rm CZ}=\pi/4|J_{12}|$. The two times match when $\Delta_n=2\sqrt{p\tilde{g}_{\parallel n}^{(1)}\tilde{g}_{\parallel n}^{(2)}}$. 

Sources of error for the CZ gate include the decay of the resonator photon that gives rise to a global dephasing channel for the two-qubit system (measurement induced dephasing), and the dephasing rate $\gamma_\phi$ due to residual noise on the individual qubits \cite{Royer2017,Harvey2018,Schuetz2017,Schuetz2019,Bottcher2021}. The fidelity of the two-qubit CZ gate was calculated in Refs.~\cite{Royer2017, Harvey2018, Schuetz2019}, taking these processes into account. When $\tilde{g}_{\parallel n}^{(1)}=\tilde{g}_{\parallel n}^{(2)}=\tilde{g}_{\parallel n}$ and the damping is small, the gate error is $1-\overline{\fid}_{\rm CZ}=\tfrac{4}{5}[\gamma_\phi +\keff(\tilde{g}_{\parallel n}/\Delta_n)^2]t_{\rm CZ}\sim(2\sqrt{p}\gamma_\phi+\keff/(2\sqrt{p}))/\tilde{g}_{\parallel n}$ \cite{Harvey2018}, where $\keff=\kr[1+2n_{th}(\omega_n)]$ is the effective resonator damping rate that accounts for the finite thermal occupation of the resonator mode \cite{Schuetz2019}. Thus a larger resonator cycle count $p$ can improve the fidelity, optimal at $p^\ast=\max(1,\,[\keff/4\gamma_\phi])$, where $[\cdot]$ is the integer closest to a number. This yields the minimal error \footnote{\mf{In the dephasing model of Ref. \cite{Schuetz2019} the two-qubit CZ gate fidelity decays linearly with time at small time. Different dephasing models can change the exponent in Eq.~(\ref{fidcz}) but this expression still generally provides a valid estimate for the gate error.}}:
\begin{equation}\label{fidcz}
1-\overline{\fid}_{\rm CZ}^\ast\sim2\max\left(\gamma_\phi\,,\sqrt{\keff\gamma_\phi}\right)/\tilde{g}_{\parallel n}\,.
\end{equation} 
We emphasize that close to zero electric field $\ef_y$, the dipolar couplings vanish, so that the dephasing rate due to quasistatic charge noise $\gamma_\phi^{\rm st}=\tfrac{\mu_B B}{\hbar}\gamma_{\parallel}\sigma_V^2$ is second-order in $\sigma_V$. Moreover, a sufficiently large $p$ allows for the integration of the entangling gate into a noise decoupling pulse sequence \cite{Harvey2018}. Indeed, since the longitudinal coupling term commutes with the qubit Hamiltonian, a spin echo protocol can make the CZ gate even more robust to quasistatic noise and decrease the effective dephasing rate $\gamma_\phi$ \cite{Harvey2018, Schuetz2019}. With the numerically computed curvature parameter shown in Fig. \ref{fig:longitudinal}, we estimate $\tilde{g}_{\parallel n}/2\pi\sim10\unit{MHz}$ with driving $V_d\sim2\unit{mV}$ in a magnetic field $B=2\unit{T}$ along $y$ ($\omega_L/2\pi\approx6\unit{GHz}$). Along with rates $\gamma_\phi\sim 50\unit{kHz}$~\cite{Piot2022} and $\keff/2\pi\sim0.5\unit{MHz}$ ($p^\ast=16$) the CZ gate error is $1-\overline{\fid}_{\rm CZ}^\ast\sim 1\%$, which could be improved in the near term by reducing the loss rate of the resonator microwave photons and by even further mitigating the dephasing noise.

Going back to the formal equivalence between the first and the second-order longitudinal couplings, we point out that Eq.~\eqref{fidcz} can also be applied to the first-order longitudinal interaction by replacing $\tilde{g}_{\parallel n}\rightarrow 2g_\parallel$, $\Delta_n\rightarrow\omega_0$ \cite{Schuetz2019}, and by properly adapting the dephasing rate parameter. In the quasistatic case, the latter reads $\gamma_\phi^{\rm st}=\tfrac{\mu_B B}{\hbar}\beta_{\parallel}\sigma_V$, which is now first-order in $\sigma_V$.

Another interesting application of the longitudinal coupling is readout. Compared to dispersive readout, performed by detuning a transversely coupled qubit-cavity system, longitudinal readout yields a much larger signal to noise ratio for similar parameters \cite{Didier2015,Blais2020}. To perform longitudinal readout, the longitudinal coupling is driven in resonance with the resonator ($\omega_d=\omega_n$), which for a single-hole qubit, can be performed with time-dependent gate voltages as above. 
The measurement is then performed through homodyne detection and is non-demolition. Furthermore, unlike with dispersive measurement, Purcell relaxation is not relevant since the longitudinal interaction commutes with the qubit Hamiltonian. 

We finally point out that the second-order longitudinal interaction described in this section also has a transverse second-order counterpart that gives rise to a two-photon term in the Hamiltonian, $\mathcal{H}=\hbar\tilde{g}_\perp\sigma_\perp[a^2+(a^\dagger)^2]$, where
\begin{equation}
\tilde{g}_{\perp}=V_{\rm zpf}^2 \frac{\mu_BB}{2\hbar}\gamma_{\perp}\,,~~\gamma_{\perp}=\left|\left(\gt''(\vg)\cdot\mathbf{b}\right)\times\mathbf{n}\right|\,.
\end{equation}
This allows for resonant two-photon processes relevant for the generation of non-classical states of radiation \cite{Blais2021}. The analysis of these processes is beyond the scope of the article. The second-order longitudinal interaction $\gamma_\parallel$ also gives rise to a dispersive shift of the spin Larmor frequency of order $\mathcal O(V_{\rm zpf}^2)$~\cite{Bottcher2021}, which is therefore negligible when $V_{\rm zpf}\ll V_d$.

\subsection{Numerical simulation of a realistic device}\label{sec:realistic}

In order to strengthen the conclusions of this work, we show that the trends highlighted in Figs. 2 and 3 of the Main Text can be reproduced in a more realistic setup with non-homogeneous lateral and vertical electric fields (hence no bias with strict inversion symmetry). 

\begin{figure}[h]
\centering
\includegraphics[width=.75\textwidth]{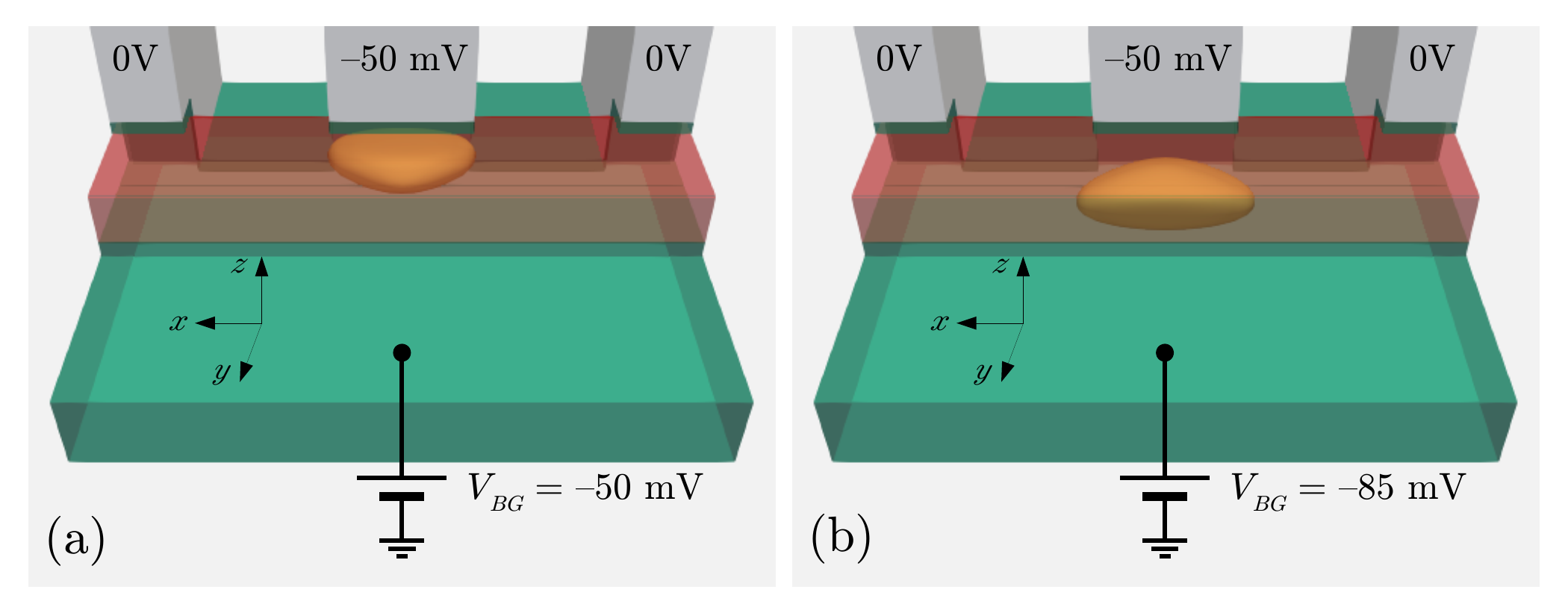}
\caption{The simulated device. The silicon channel is in red, the silicon oxide in green and the gates in gray. The whole device is embedded in Si$_3$N$_4$ (not shown). The dot is controlled by the central front gate at potential $\vg=-50\unit{mV}$, while the side gates are grounded. The substrate below is used as a back gate at potential $V_{BG}$. The iso-probability surface of the ground-state wave-function that encloses 90\% of the hole charge is shown at (a) $V_{BG}=-50\unit{mV}$ and (b) $V_{BG}=-85\unit{mV}$.}
\label{fig:device}
\end{figure}

The device we consider is similar to the one studied in Ref.~\cite{Venitucci2018} (see Fig.~\ref{fig:device}). It consists in a rectangular, $[110]$ oriented silicon nanowire with $(001)\times(\bar{1}10)$ facets and sides $L_z=15\unit{nm}$ and $L_y=40\unit{nm}$. This nanowire lies on a $25\unit{nm}$ thick buried oxide and a substrate used as a back gate at potential $V_{BG}$. A central $L_x=40\unit{nm}$ long front gate overlaps the channel by $S_y=10\unit{nm}$. It is insulated from the channel by a $4\unit{nm}$ thick SiO$_2$ layer, and biased at $\vg=-50\unit{mV}$. 

Two other front gates laid $40\unit{nm}$ on the left and right mimic other qubits or ``access'' gates to reservoirs. They are grounded in the present study. The whole device is embedded in Si$_3$N$_4$. The central conductor of the microwave resonator is connected to the central front gate.

Poisson's equation for the electrostatic potential landscape in the device is solved with a finite-volumes method as a function of the back gate voltage $V_{BG}$. The wave-function of the single hole trapped under the central front gate is then computed with a 6 bands $\mathbf{k}\cdot\mathbf{p}$ model (including the nearby split-off bands) discretized using a finite-differences scheme. The $\gt$-matrices and their derivatives are finally calculated along the lines of Ref.~\cite{Venitucci2018}.  

\begin{figure}[h!]
\centering
  \includegraphics[width=0.32\textwidth]{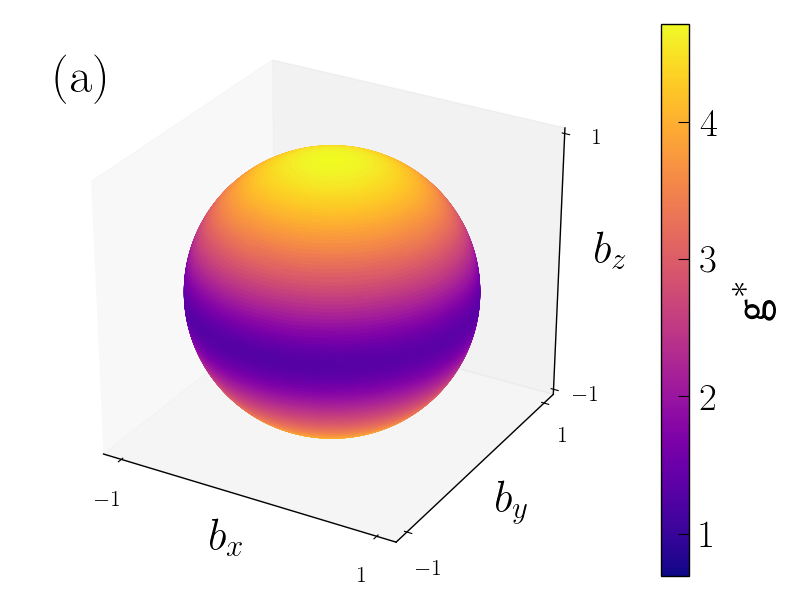}
  \includegraphics[width=0.32\textwidth]{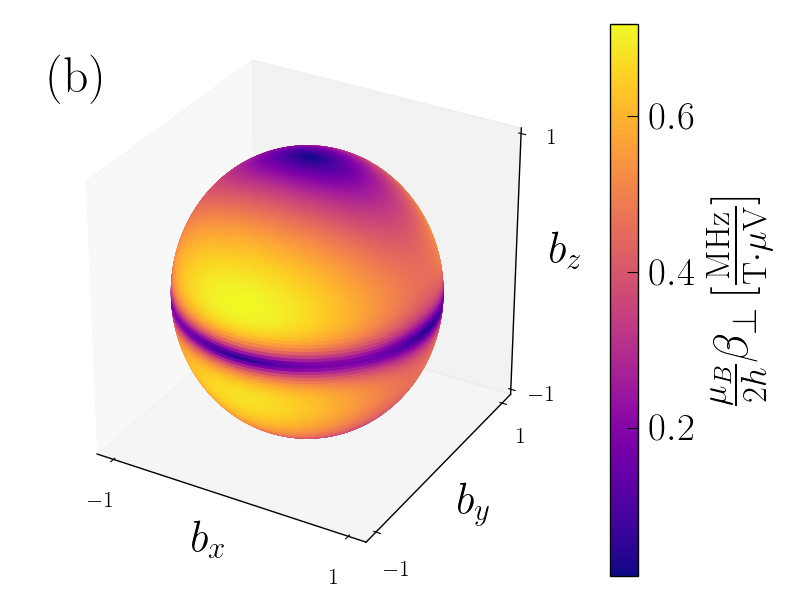}
  \includegraphics[width=0.32\textwidth]{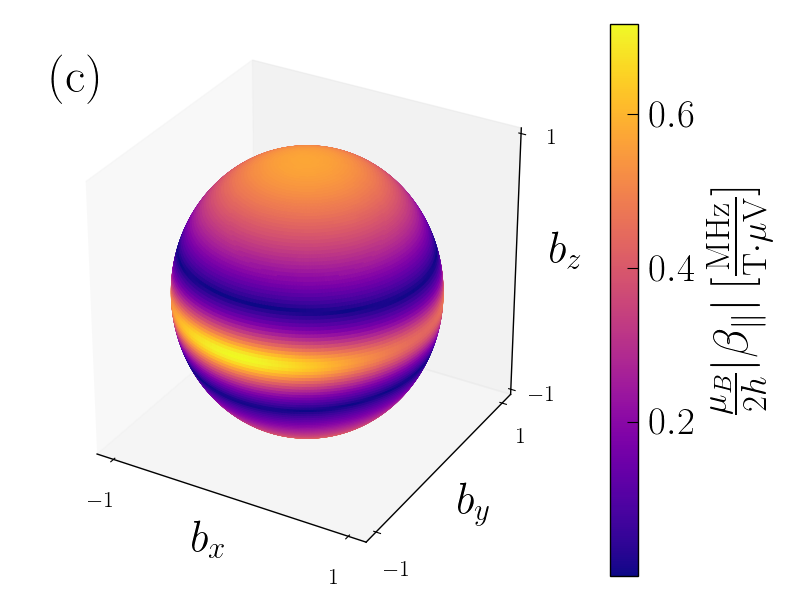}
  \includegraphics[width=0.32\textwidth]{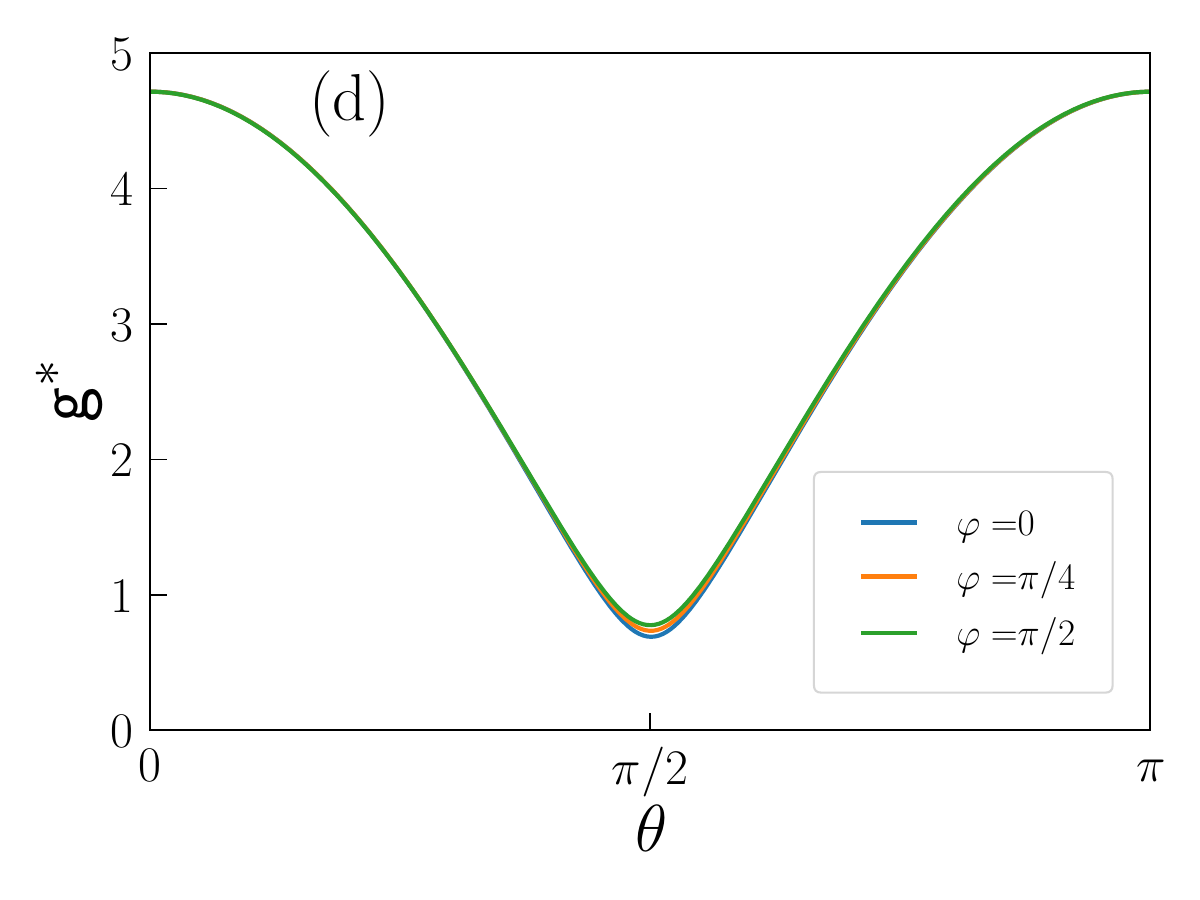}
  \includegraphics[width=0.32\textwidth]{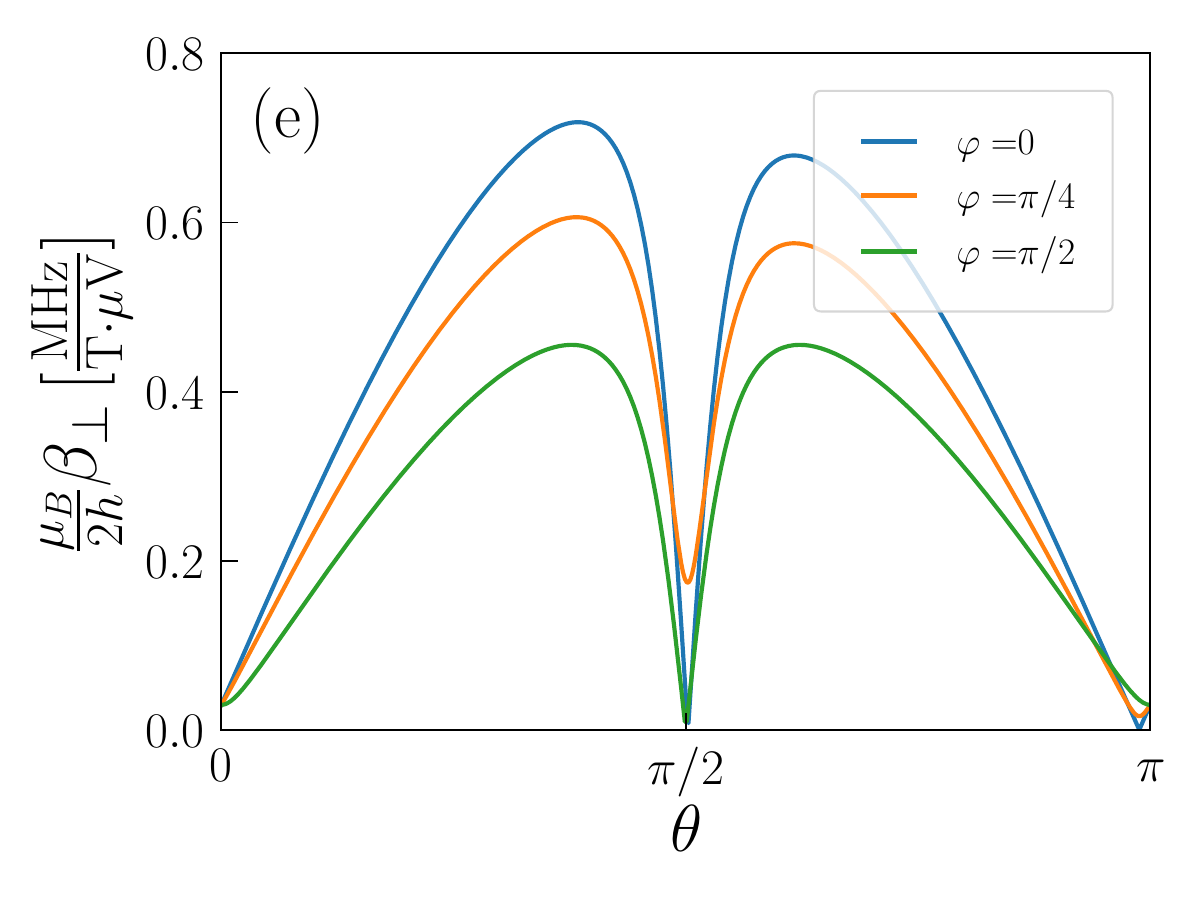}
  \includegraphics[width=0.32\textwidth]{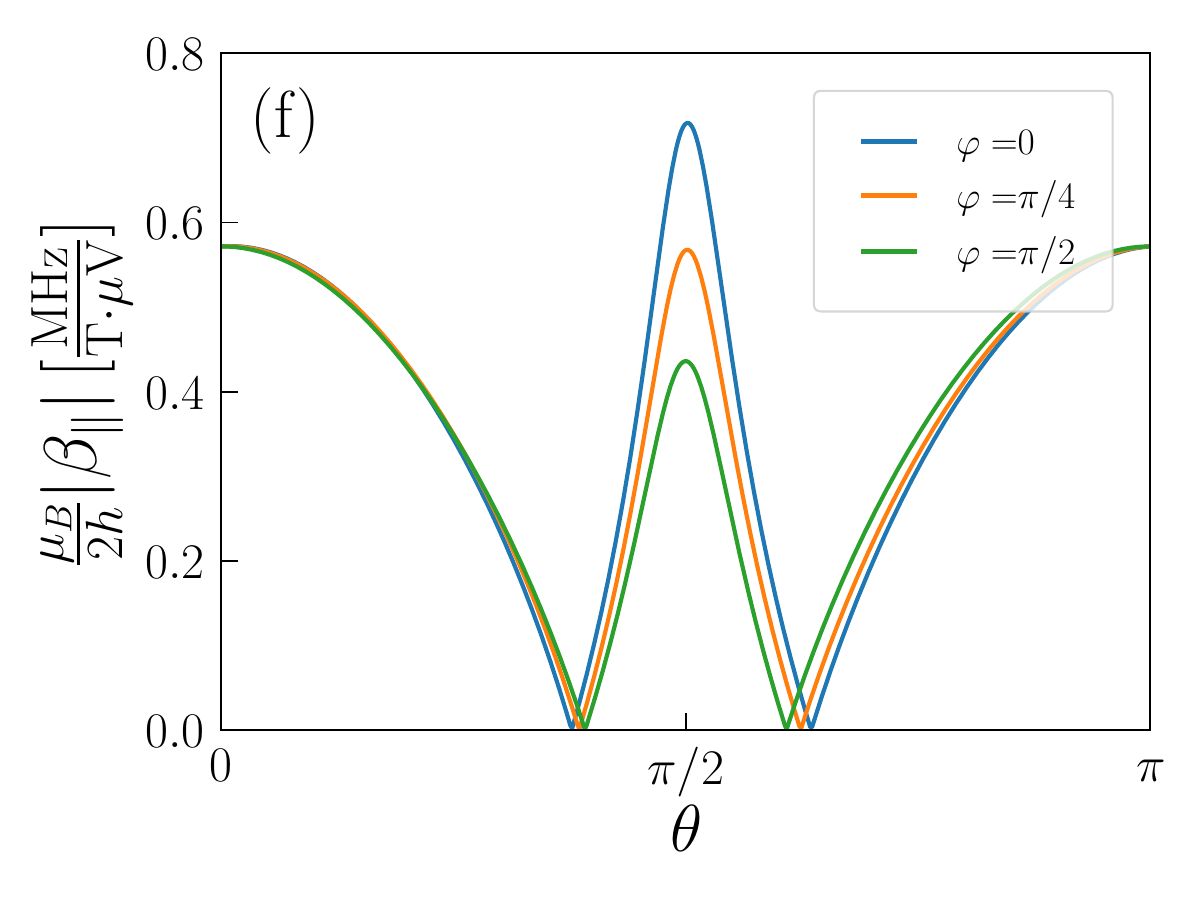}
  \includegraphics[width=0.9\textwidth]{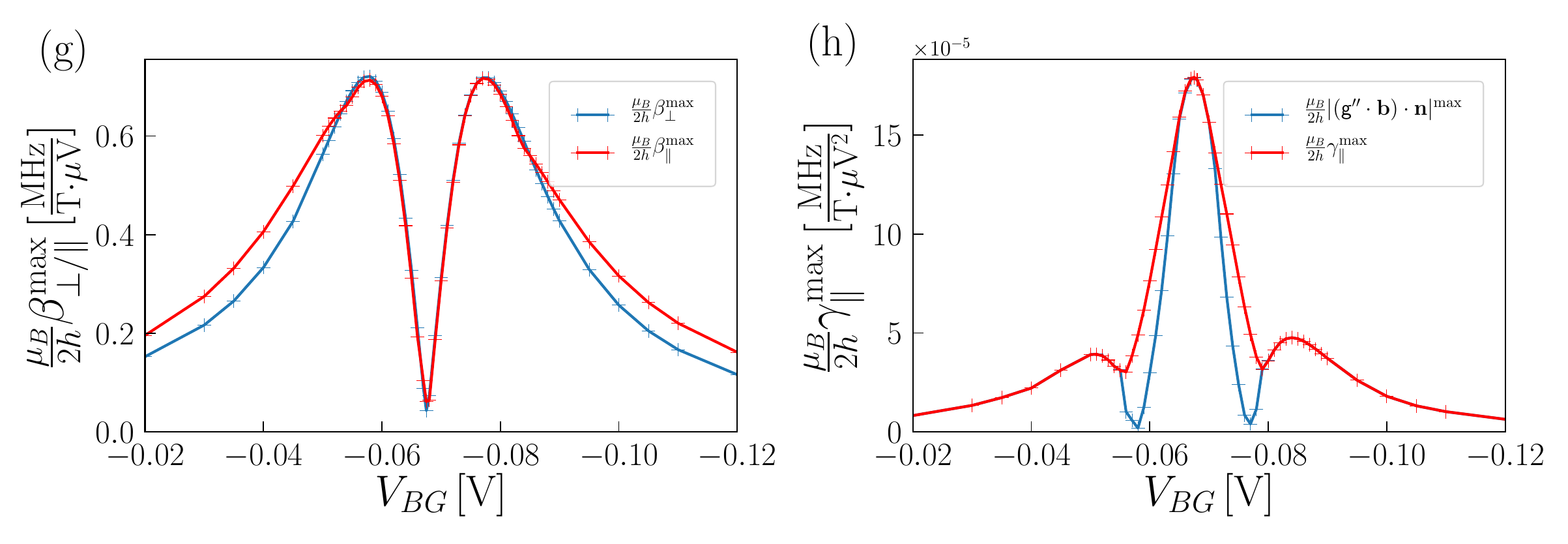}
\caption{Maps of (a) the effective $\gt$-factor and (b), (c) the perpendicular $\beta_{\perp}$ and parallel $|\beta_{\parallel}|$ susceptibilities as a function of the magnetic field orientation $\mathbf{b}=\mathbf{B}/|\mathbf{B}|$, calculated in the device of Fig.~\ref{fig:device} at $V_{BG}=-77\unit{mV}$. (d), (e), (f) Plots of the same quantities at constant azimuthal angle $\varphi$ (measured from the $y$ axis) as a function of the polar angle $\theta$ from the $z$ axis ($\varphi=0$ is hence the $yz$ plane and $\varphi=\pi/2$ the $xz$ plane). (g) Maximum parallel and perpendicular susceptibilities $\beta_{\parallel}^{\max}$ (red) and $\beta_{\perp}^{\max}$ (blue) as a function of the back gate voltage $V_{BG}$. (h) Curvature parameter $\gamma_{\parallel}^{\max}$ (red) as a function of the back gate voltage $V_{BG}$ ($\mathbf{B}\parallel\mathbf{y}$). The contribution from the first term of Eq.~\eqref{eq:curvature2} is also plotted in blue.}
\label{fig:simulation}
\end{figure}

The $g$-factor, perpendicular susceptibility $\beta_\perp$, and parallel susceptibility $\beta_\parallel$ are plotted as a function of the orientation of the magnetic field $\mathbf b$ in Figs.~\ref{fig:simulation}a-f, at back gate voltage $V_{BG}=-77\unit{mV}$. They show the same trends as Fig.~2 of the Main Text. The maximal susceptibilities  $\beta_{\parallel}^{\max}$ and $\beta_{\perp}^{\max}$ (with respect to the magnetic field orientation) are plotted as a function of $V_{BG}$ in Fig.~\ref{fig:simulation}g. When sweeping the back gate voltage from positive to negative values, the hole moves from the right facet of the channel (under the front gate) to the left facet (opposite to the front gate)~\cite{Venitucci2018}. At $V_{BG}\approx-68\unit{mV}$, the hole wave-function spans the whole width of the channel. This is the situation closest to $\ef_y=0$ in the model device of the Main Text, although there is no strict inversion symmetry in the confinement potential at this bias. Yet $\beta_{\parallel}^{\max}$ and $\beta_{\perp}^{\max}$ do vanish as the orientation of $\gt'\cdot\mathbf{b}$ reverses when the hole moves from one facet to the other. The longitudinal curvature parameter $\gamma_{\parallel}^{\max}$ (for $\mathbf{B}\parallel\mathbf{y}$) is also optimal at that back gate voltage (see Fig.~\ref{fig:simulation}h). Therefore, the device of Fig.~\ref{fig:device} behaves qualitatively as the model device of the Main Text.

From a quantitative point of view, the optimal susceptibilities $\beta_\parallel^\ast$, $\beta_\perp^\ast$ and $\gamma_\parallel^\ast$ are of the same order of magnitude as in the model device. Voltage drops in the oxides and other materials around the channel however reduce the $\beta$'s by up to 40\%. This is (over-)compensated for the $\gamma$'s by enhanced non-linearities. The smaller the overlap between the gate and the channel, the larger the ratio between the $y$ and $z$ components of the electric field (the spin-photon couplings are significantly stronger for a $S_y=10\unit{nm}$ than for a $S_y=20\unit{nm}$ overlap, by up to a factor two for $\gamma_\parallel$). The present device might be further optimized to limit voltage drops and enhance the electric field along $y$, with, e.g., the introduction of ``face-to-face'' gate layouts~\cite{Roche2012}, at the expense of a more complex manufacturing.

\end{document}